\documentclass[acmsmall,authorversion,nonacm]{acmart}

\usepackage[utf8]{inputenc}
\usepackage[T1]{fontenc}
\usepackage{hyperref}
\usepackage{url}
\usepackage{amsfonts}
\usepackage{mathtools}
\usepackage{xcolor}
\usepackage{tabulary}
\usepackage{subcaption}
\usepackage[normalem]{ulem}
\newcommand{\xhdr}[1]{\vspace{0.2mm}\noindent{{\bf #1.}}}

\begin{document}

\title{(De)Noise: Moderating the Inconsistency Between Human Decision-Makers}

\author{Nina Grgić-Hlača}
\affiliation{%
  \institution{Max Planck Institute for Software Systems, Max Planck Institute for Research on Collective Goods}
  \country{Germany}}
\email{nghlaca@mpi-sws.org}

\author{Junaid Ali}
\affiliation{%
  \institution{Max Planck Institute for Software Systems}
  \country{Germany}}
\email{junaid@mpi-sws.org}

\author{Krishna P. Gummadi}
\affiliation{%
  \institution{Max Planck Institute for Software Systems}
  \country{Germany}}
\email{gummadi@mpi-sws.org}

\author{Jennifer Wortman Vaughan}
\affiliation{%
  \institution{Microsoft Research}
  \country{United States}}
\email{jenn@microsoft.com}

\thanks{Author contribution statement: All authors contributed to conceiving the idea, designing the experiments and interpreting the result. Nina Grgić-Hlača, Junaid Ali and Jennifer Wortman Vaughan contributed to writing the paper. Nina Grgić-Hlača and Junaid Ali gathered the experimental data. Junaid Ali took the lead in implementing the survey instruments and decision aids, and in conducting the exploratory analyses. Nina Grgić-Hlača took the lead in drafting the paper, and in conducting the statistical analyses.}
\thanks{Funding statement: This research was supported in part by a European Research Council (ERC) Advanced Grant for the project ``Foundations for Fair Social Computing'', funded under the European Union’s Horizon 2020 Framework Programme (grant agreement no. 789373).}
\thanks{Acknowledgements: We thank all the participants in our study. We are grateful to the CSCW reviewers for their constructive feedback. We thank Neda Foroutan for participating in the early stages of implementing the decision aids.}

\begin{abstract}
  Prior research in psychology has found that people’s decisions are often inconsistent \cite[Kahneman et al.,][]{kahneman2016noise,kahneman2021noise}. An individual’s decisions vary across time, and decisions vary even more across people. Inconsistencies have been identified not only in subjective matters, like matters of taste, but also in settings one might expect to be more objective, such as sentencing, job performance evaluations, or real estate appraisals. In our study, we explore whether algorithmic decision aids can be used to moderate the degree of inconsistency in human decision-making in the context of real estate appraisal. In a large-scale human-subject experiment, we study how different forms of algorithmic assistance influence the way that people review and update their estimates of real estate prices. We find that both (i) asking respondents to review their estimates in a series of algorithmically chosen \emph{pairwise comparisons} and (ii) providing respondents with \emph{traditional machine advice} are effective strategies for influencing human responses. Compared to simply reviewing initial estimates one by one, the aforementioned strategies lead to (i) a higher \emph{propensity to update} initial estimates, (ii) a higher \emph{accuracy} of post-review estimates, and (iii) a higher degree of \emph{consistency} between the post-review estimates of different respondents. While these effects are more pronounced with traditional machine advice, the approach of reviewing algorithmically chosen pairs can be implemented in a wider range of settings, since it does not require access to ground truth data.
\end{abstract}



\maketitle

\section{Introduction} \label{sec:intro}

\begin{quote}
    ``… humans are unreliable. Judgments can vary a great deal from one individual to the next, even when people are in the same role and supposedly following the same guidelines.'' 
\end{quote}
\begin{flushright}
    \citet{kahneman2016noise}
\end{flushright}

Presented with identical information, the same person might make different decisions at different points in time, and the decisions of different people are likely to vary even more \cite{kahneman2016noise, kahneman2021noise}. Such inconsistencies between decision-makers have been identified in numerous settings including sentencing \cite{anderson1999measuring}, job performance evaluations \cite{taylor1974capturing}, real estate appraisals \cite{adair1996analysis}, and---especially close to the research community---conference reviewing \cite{lawrence2021neurips, beygelzimer2021neurips, tran2020open, cortes2021inconsistency, bruckman2017cscw}.

In certain settings, variation in people’s decisions is indispensable; it may contain invaluable information that reflects the variation in people’s background knowledge, political or moral stances, life experiences, and other factors \cite{wanous1986solution, roberge2010recognizing, welinder2010multidimensional}. However, in other settings, consistency may be considered normatively desirable instead. For instance, \citet{kahneman2016noise} argue that organizations such as credit-rating and insurance agencies expect that, regardless of the particular professional handling each case, ``identical cases should be treated similarly, if not identically’’---a notion in line with that of ``individual fairness'' in the algorithmic fairness literature \cite{dwork2012fairness}. In conference reviewing, inconsistencies between different groups of reviewers have raised concerns about the peer-review process in the scientific community \cite{lawrence2021neurips, beygelzimer2021neurips, tran2020open, cortes2021inconsistency, bruckman2017cscw}. In the organizational justice literature, consistency of decisions is recognized as an important aspect of procedural justice \cite{lee2019procedural, leventhal1980should}. 

In this work, we focus on such settings where consistency between decision makers might be deemed desirable, and study how the degree of inconsistency of human decisions could be moderated with algorithmic assistance. This has immediate implications for the development of algorithmic assistance to support cooperative work by enabling the distribution of decision-making tasks amongst multiple decision-makers, without sacrificing consistency. Specifically, we leverage prior work in psychology and HCI (reviewed in Section \ref{sec:related_work}) to develop a set of algorithmic decision aids which may influence the degree of inconsistency of human decisions, and we rigorously experimentally evaluate the effect of these decision aids on human decisions.

\begin{figure}[t]
    \centering
    \includegraphics[width=0.8\textwidth,trim={0cm 7cm 0cm 3cm},clip]{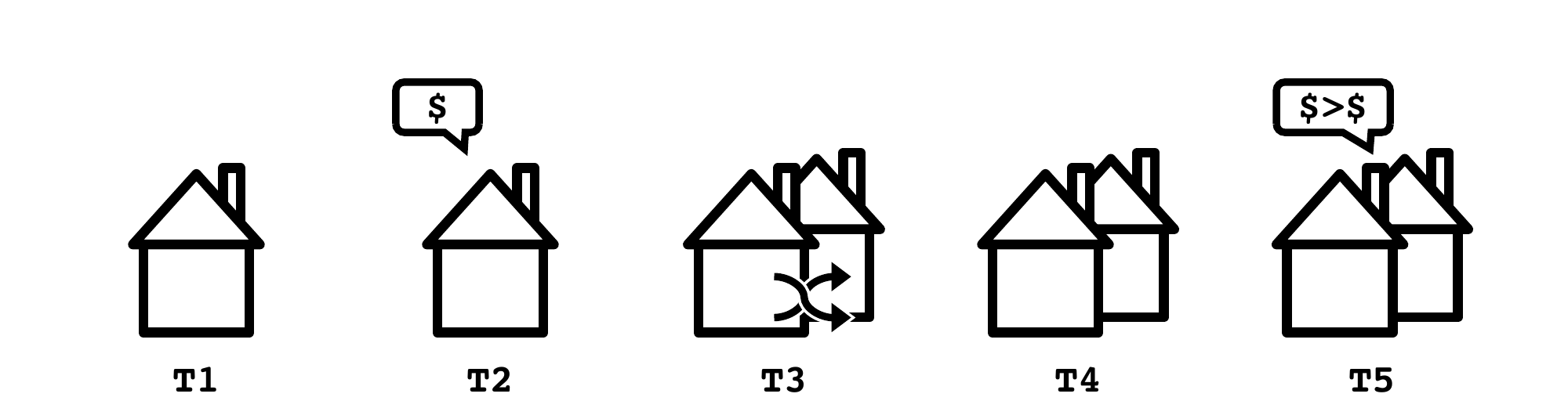}
    \caption{Graphical overview of experimental conditions T1-T5. In T1 and T2, respondents review their decisions one-by-one, while in T3-T5 they review decisions in randomly (T3) or meaningfully (T4 and T5) selected pairs. In T2 and T5 respondents are additionally provided with (different kinds of) explicit machine advice.}
    \label{fig:experimental_conditions}
\end{figure}

\xhdr{Experimental Design}
In a large-scale human-subject experiment ($N=643$), we explored five different approaches to moderating human inconsistency, summarized in Figure \ref{fig:experimental_conditions}. We focused on the task of estimating real estate prices, using a real estate price dataset studied by \citet{poursabzi2021manipulating}. As a baseline (T1), we measured how people's decisions are affected by having the opportunity to review them one-by-one. We compared the effectiveness of this simple reviewing procedure with a series of more sophisticated interventions. 

Inspired by the growing popularity of machine-assisted decision-making in the judge-advisor system (JAS) paradigm \cite{bonaccio2006advice}, we studied the effect of providing machine advice (T2). Leveraging prior research in psychology which found that people might find it easier to make pairwise comparisons than absolute judgments \cite{miller1956magical, stewart2005absolute}, we also investigated the effectiveness of reviewing past judgments as a series of pairwise comparisons. We studied the effectiveness of asking respondents to review randomly selected pairs of apartments (T3), or pairs of apartments whose price estimates were inconsistent with most people's pairwise valuations---with (T5) and without (T4) explicitly informing participants about this inconsistency.

\xhdr{Contributions}
We find that both providing traditional \emph{machine advice} (T2) and asking respondents to review their past decisions as a series of meaningfully selected \emph{pairwise comparisons} (T4 and T5) have a large and statistically significant effect on people's decisions. 

Namely, in our pre-registered confirmatory analysis we find that compared to reviewing past decisions one-by-one (T1), our interventions (T2, T4 and T5) lead to:
\begin{itemize}
    \item a higher \emph{propensity of updating initial decisions}
    \item a higher \emph{accuracy} of decisions after the review phase
    \item a higher degree of \emph{consistency} amongst the post-review decisions of different respondents
\end{itemize}

In the paper's appendix, we also conduct an exploratory analysis of the effects of our interventions on different measures of accuracy and consistency, and report a series of other descriptive statistics.

\xhdr{Takeaways}
Prior work on machine-assisted decision-making (as reviewed in Section \ref{sec:related_work}) has showcased that traditional machine advice (T2) can influence people's decisions \cite{grgic2019human} and---if the decision aid is sufficiently accurate---improve people's decision accuracy \cite{grgic2022taking, yin2019understanding}. We demonstrate that traditional machine advice can also increase the consistency between people's decisions. 

We additionally go beyond studying traditional machine advice and demonstrate the effectiveness of novel alternative strategies.  Asking people to review their decisions in a series of meaningfully selected pairwise comparisons (T4 and T5) significantly increases the accuracy and consistency of their decisions. Unlike traditional machine advice (T2), this approach does not require access to ground truth data---in our case, apartment prices. It requires only data about people's comparative valuations of apartments, which is used to identify pairs where a person's relative ordering of apartments does not align with the majority's estimates. Hence, it is also applicable in settings where ground truth may be costly to obtain or difficult to define. We note that in settings where there is no clearly defined notion of ground truth, such as in subjective tasks, one cannot make meaningful claims about the effects of algorithmic assistance on the \emph{accuracy} of people's decisions. Still, such a decision aid could be used to increase inter-annotator \emph{consistency}, if it is deemed normatively desirable to do so.

Interestingly, when asking people to review meaningfully selected pairs of estimates (T4), we observe no significant effect of explicitly informing people that their relative ordering of apartments did not align with the majority's comparative valuations (T5). Hence, we identified an effective strategy for increasing the accuracy and consistency of people's responses that is applicable even in scenarios where it is normatively undesirable to explicitly steer people towards specific decisions. 

Through our exploratory analysis in the appendix, we see that these results appear robust to alternative notions of accuracy and consistency.
\section{Background} \label{sec:related_work}

In this section, we provide an overview of related work and position our contributions with respect to this body of research. We commence by reviewing the literature on machine-assisted decision-making. Next, we review prior research on the inconsistency of human decisions, discussing different notions of inconsistency and different approaches to reducing inconsistency. Finally, we review prior research in social psychology to form hypotheses about people's reactions to receiving feedback about their inconsistency.

\subsection{Machine-Assisted Decision-Making}
With the increasing popularity of algorithmic decision aids, much research has studied how algorithmic advice shapes human decisions. At a high level, we can discuss the findings of this research in terms of the \emph{factors} that were identified to influence people's advice taking behavior, and the \emph{measures} used to evaluate the effects of machine advice on people's decisions.

Research has identified a plethora of factors that influence people's advice taking behavior. The likelihood of accepting advice varies based on the advisor's identity, with a preference for human guidance in certain scenarios \cite{burton2020systematic, dietvorst2015algorithm, dietvorst2018overcoming}, and a preference for algorithmic advice in other settings \cite{logg2017theory, logg2019algorithm}. Receptiveness to machine advice depends on the algorithm's errors. People are more likely to take advice from algorithms that are stated or observed to exhibit a higher predictive accuracy \cite{yin2019understanding}, and that make errors more similar to typical human errors \cite{grgic2022taking}. People are also more likely to take advice from decision aids that are explainable \cite{chen2023understanding, poursabzi2021manipulating, wang2021explanations}. On the other hand, strong graphical warnings about algorithmic decision aids may lower the influence of their advice \cite{engel2021machine}. Advice taking also depends on the specific advice that is provided; for instance, in the legal domain, people are more likely to take advice to grant a defendant bail than advice to deny bail \cite{grgic2019human}. 

Prior work has studied how machine advice affects human decisions with respect to a variety of measures. As a first step, most studies measure people's likelihood of taking machine advice, be it in terms of the overall advice taking propensity \cite{yin2019understanding}, the propensity to take advice pointing towards a specific decision \cite{grgic2019human}, or the propensity to take (in)correct advice \cite{poursabzi2021manipulating}. Some studies have gone beyond measuring people's likelihood of taking advice, and measured the effects of machine advice on the quality of people's decisions. For instance, in settings where one can define the notion of correct predictions and ground truth labels, prior work has measured whether machine advice increases the alignment between people's decisions and ground truth labels---that is, the accuracy of people's decisions \cite{zhang2020effect, grgic2022taking, poursabzi2021manipulating}. In settings where decisions have important societal implications, prior work has also studied the effects of algorithmic assistance on the fairness of people's decisions \cite{green2019disparate, green2019principles, yang2024fair}. However, little prior work considered the effects of algorithmic assistance on the \emph{consistency} of human decisions.

We contribute to research on machine-assisted decision-making by studying how different algorithmic decision aids impact human decisions with respect to two measures established in prior work---people's propensity to take advice and the accuracy of people's decisions---as well as a novel measure: the consistency between people's decisions.

\subsection{The (In)Consistency of Human Decisions}

\xhdr{Notions of Inconsistency} Much prior work has studied the (in)consistency of human decisions. A bulk of research has documented the \emph{inconsistency between decisions of different decision-makers}~\cite{kahneman2016noise, kahneman2021noise}---that is, a lack of \emph{inter}-annotator consistency---in tasks as diverse as sentencing \cite{anderson1999measuring}, evaluating job performance \cite{taylor1974capturing}, estimating real estate prices \cite{adair1996analysis}, and reviewing submissions to top-tier CS conferences, such as NeurIPS \cite{beygelzimer2021neurips, cortes2021inconsistency, lawrence2021neurips}, CSCW \cite{bruckman2017cscw} and ICLR \cite{tran2020open}. We contribute to this line of research by exploring if algorithms can be used to support cooperative work in such settings where consistency is deemed to be desirable. Namely, we propose methods for alleviating the inconsistency between the decisions of different decision-makers for the same set of inputs.

Much research has also studied the consistency of individual decision makers, or \emph{intra}-annotator consistency. Cognitive biases such as dynamic inconsistency and hyperbolic discounting are known to result in the \emph{inconsistency of an individual's decisions across time} \cite{loewenstein1992anomalies, thaler1981some}. Individual's judgments are found to substantially vary across time in various settings~\cite{kahneman2016noise}: pathologist's biopsy assessments of the same sample at different points in time were found to exhibit a correlation of only 0.61 \cite{einhorn1974expert}; experts' estimates of the amount of time required to complete the same software development task were found to vary by 71\% \cite{grimstad2007inconsistency}. 

Prior work has also documented the \emph{inconsistency of an individual's judgments across inputs}. A particularly well-studied aspect of this problem is the inconsistency of people's pairwise preferences. Decades of research in this area have led to the development of numerous methods for identifying, measuring, and reducing the inconsistency of human pairwise preferences \cite{koczkodaj1993new, brunelli2015axiomatic, abel2018inconsistency}, as well as a plethora of approaches to the difficult task of learning human pairwise preferences \cite{hullermeier2008label, furnkranz2003pairwise, cheng2010predicting}, which we consult when developing our decision aids in Section \ref{subsec:decision_aids}.

\xhdr{Human Heuristics for Reducing Inconsistency}
Many decision-making settings require people to make decisions on a case-by-case basis: granting or denying loans, making bail decisions, reviewing papers, etc. However, prior research in psychology has documented that people might find it easier to make \emph{comparative} judgments than absolute ones in various contexts \cite{stewart2005absolute, miller1956magical}. Pairwise comparisons are also used to assist people with developing and refining their beliefs \cite{lee2019webuildai}. Hence, it is not surprising that people often rely on comparative judgments to assist them with making case-by-case decisions. For instance, the analytical hierarchy process that is widely used to assist with making complex decisions in domains ranging from governance to engineering relies on comparative judgments at its core \cite{forman2001analytic, vaidya2006analytic}. Research on crowdsourcing has also proposed eliciting respondents' relative pairwise labels, rather than absolute ones, as a strategy for improving response quality \cite{chen2013pairwise,narimanzadeh2023crowdsourcing,sunahase2017pairwise}.

To illustrate how one may leverage comparative judgments to assist with absolute judgments, let us consider the task of grading papers. After (i) assigning initial grades to a set of papers, one might (ii) compare pairs (or larger subsets) of papers to identify mutually inconsistent decisions, in order to (iii) revise the final grades. In T4 and T5, we develop a decision aid that identifies pairs of decisions that are inconsistent with the majority's comparative valuations, hence providing a tool for automating step (ii) in the above-described procedure.

\xhdr{Reducing Inconsistency with Algorithmic Assistance}
Kahneman at al. \cite{kahneman2016noise, kahneman2021noise} extensively study the problem of noise in human judgments. In \citet{kahneman2016noise}, they discuss several approaches to reducing inconsistency in human decisions, proposing interventions of varying strengths. The first and most radical proposal is to replace human decision makers with algorithms. Still, they highlight the need for people to retain ultimate control. Hence, as the second and weaker proposal, they propose the use of algorithmic decision aids to assist human decisions. Depending on the estimated accuracy of human and algorithmic decisions and the normative importance of accuracy, they highlight the possibility of advising against overruling algorithmic predictions. The third and weakest intervention is ensuring that decision-makers use similar procedures to gather and integrate information, and to translate this information into a decision. 

The first two proposals rely on algorithmic assistance to reduce human inconsistency. The underlying idea is to use algorithms to \emph{predict correct decisions}, which would steer (second proposal) or even replace (first proposal) human decisions, thereby making them more consistent. In this paper, we study the effects of the second proposal in treatment T2. However, we also study an alternative approach in treatments T4 and T5: using algorithms to \emph{identify inconsistencies} in human decisions, and to help people reduce their inconsistency by themselves.

\subsection{Anticipated Reactions to Feedback about Inconsistency}
Algorithmic decision aids have proven to be effective in a plethora of settings. Here we review literature in social psychology that may help us form hypotheses about the effectiveness of algorithmic decision aids for the task of reducing inconsistency in human decisions, and guide the design of our decision aids. Specifically, we leverage prior work in social psychology to anticipate how people may react to being provided with feedback about their inconsistency.

\citet{kahneman2016noise} argue that inconsistency is undesirable in a variety of settings. If our respondents share this view, they might perceive feedback about their inconsistency as negative feedback. Prior work in organizational psychology has shown that people may not react positively to negative feedback about their performance. Negative feedback is not perceived as useful, results in negative reactions, and is not associated with a recipient’s willingness to change their behavior \cite{steelman2004moderators}. It is also found to evoke defensiveness and denial \cite{london2003job}. The main strategy employees use to reduce the impact of such negative feedback is to reject it \cite{ilgen1979consequences}. To mitigate these effects, we will avoid framing the machine advice as negative feedback, and utilize strategies for softening the blow proposed by \citet{steelman2004moderators}: providing high quality feedback delivered in a considerate manner.

\citet{kahneman2016noise} also show that people tend to vastly underestimate the degree of inconsistency in human decision-making. Hence, feedback about inconsistency may conflict with people’s beliefs. Much prior work in psychology has found that people resist evidence that is contradictory to their preconceptions \cite{allport1954nature}. Two psychological concepts that are particularly relevant for predicting how people will react to conflicting information are cognitive dissonance \cite{harmon2019introduction, festinger1962theory} and biased assimilation \cite{lord1979biased} or disconfirmation bias \cite{edwards1996disconfirmation}. Both lines of research point to the same conclusion: due to overestimating their consistency, people may discount or reject the decision aids' feedback about their inconsistency. We attempt to mitigate this effect by familiarizing people with their lack of expertise with the task at hand: at the beginning of the experiment, participants complete a tutorial where they can observe the (in)accuracy of their real-estate price estimates.
\section{Methodology} \label{sec:methodology}

In this section,  we describe our methodology, including  the details of the experimental design (Section \ref{subsec:experimental_design}), the stimulus materials (Section \ref{subsec:stimulus_material}), the data collection procedure (Section \ref{subsec:data_collection}), and the procedure of developing the decision aids that were used in our experiment (Section \ref{subsec:decision_aids}).

\subsection{Experimental Design} \label{subsec:experimental_design}
In a large-scale, pre-registered human-subject experiment run on Prolific we studied how different interventions influence people's estimates of real estate prices. The interventions included asking respondents to review their initial estimates one by one (treatments T1 and T2) or as pairwise comparisons (treatments T3, T4 and T5), and providing respondents with different forms of algorithmic assistance (treatments T2, T4 and T5). Prior to conducting the human-subject experiment, we obtained the approval of the Ethical Review Board of the Faculty of Mathematics and Computer Science at Saarland University, and pre-registered our experiment on AsPredicted. The pre-registration documentation can be found on the following url: \url{https://aspredicted.org/aw2ic.pdf}.

\xhdr{Scenario}
In this work, we focused on the task of estimating real estate prices. In our experiment, we utilized a dataset of New York City real estate prices introduced by \citet{poursabzi2021manipulating}. The dataset contains information about 393 apartments located on the Upper West Side of New York City, which were listed for sale on the real estate website StreetEasy.com between 2013 and 2015. For each listing, we had access to basic information about the apartment, including the listing price, number of bedrooms, bathrooms and total number of rooms, the apartments' square footage and monthly maintenance fees, the number of days the apartment has been on the market, and the distance from the apartment to the nearest subway and school. We preprocess the data as proposed by \citet{poursabzi2021manipulating}. Specifically, we remove the apartments where the number of bedrooms is greater than the total number of rooms or where the apartment's square footage is less than 200 sqft. With this preprocessing, we were left with 387 apartments. From these, we utilized 30 apartments as stimulus material in the human-subject experiment (Section \ref{subsec:stimulus_material}), and the remaining 357 apartments for training the decision aids (Section \ref{subsec:decision_aids}).

\xhdr{Experimental Conditions} 
In each experimental condition, respondents were first asked to complete a tutorial, in order to familiarize themselves with real estate prices in New York City. Next, all respondents were asked to estimate the prices of the same 30 apartments. After gathering the respondents' initial estimates of apartment prices, we asked them to review their estimates in one of five different ways, as described below and summarized in Figure \ref{fig:experimental_conditions} and Table \ref{tab:treatments}.

\begin{table}[t]
\centering
\begin{tabulary}{\linewidth}{LCCC}
    \toprule
    & {\bf Reviewing Procedure} & {\bf Algorithmic Assistance} & {\bf Data Required} \\
    \hline
    {\bf T1} & one-by-one & none & none \\
    {\bf T2} & one-by-one & explicit advice & ground truth \\
    {\bf T3} & pairwise comparisons & none & none \\
    {\bf T4} & pairwise comparisons & implicit advice & human perceptions \\
    {\bf T5} & pairwise comparisons & implicit and explicit advice & human perceptions \\
    \bottomrule
\end{tabulary}
\caption{Overview of the characteristics of the 5 experimental conditions in our study. Reviewing Procedure: Are instances reviewed one-by-one or pairwise? Algorithmic Assistance: Do respondents have access to any form of algorithmic assistance? Data Required: Do the utilized decision aids require any type of labeled data?}
\label{tab:treatments}
\end{table}

In the control condition T1, respondents were asked to perform the simplest revision procedure. They were asked to revise their initial estimates one-by-one.
\begin{description}
    \item[T1:] Respondents were asked to review all of their estimates \emph{one-by-one}, in the same format as they originally made them: 30 apartments, one per page, shown in random order.
\end{description}

T2 was inspired by decision aids that are commonly used in the machine-assisted decision-making literature \cite{grgic2019human, yin2019understanding}. It corresponds to the standard machine-assisted decision-making setting in the judge-advisor system (JAS) paradigm \cite{bonaccio2006advice}.
\begin{description} 
    \item[T2:] Compared to T1, we manipulated the information provided in the review phase. Respondents were again asked to review all of their initial estimates one-by-one (30 apartments, one per page, shown in random order), but the apartment descriptions were accompanied by \emph{machine advice}. Specifically, respondents were shown the estimates of a linear regression model which we trained to estimate real estate prices using the dataset introduced by \citet{poursabzi2021manipulating}, as described in Section \ref{subsec:decision_aids}.
\end{description}

While T1 and T2 required respondents to review their decisions one-by-one, in T3--T5 decisions were reviewed in pairs. This pairwise revision procedure was motivated by past research in psychology \cite{stewart2005absolute, miller1956magical} and computer science \cite{narimanzadeh2023crowdsourcing}, which found that, in certain contexts, people are better at making comparative judgments than absolute ones.

One of the main difficulties in introducing a reviewing procedure based on pairwise comparisons is selecting which pairs one should review. Our set of 30 apartments results in 30 choose 2 = 435 possible pairwise comparisons. Since it is not feasible to review all of these pairs, one must select which pairs to review. A naive approach, employed in treatment T3, would consist of randomly selecting which pairs to review. An ideal approach would consist of reviewing exactly those pairs that would lead to an improvement in the quality of decisions with respect to a metric of interest. In treatments T4 and T5 we utilize algorithmic assistance to identify pairs of apartments for which people's estimates are not aligned with the majority's estimates.

\begin{description}
    \item[T3:] In T3, respondents were asked to review their decisions in a series of 15 \emph{pairwise comparisons of randomly selected pairs} of apartments. Respondents were asked to review 15 pairs of apartments in order to keep the number of decisions reviewed equal to 30 across all treatments. The same apartment may have been shown in multiple pairs. Hence, even though the 15 pairwise comparisons provided respondents with 30 opportunities to update their estimates, this does not imply that they had an opportunity to update their initial estimate for each of the 30 unique apartments.
\end{description}

Due to its simplicity and lack of an algorithmic component, we treat T3 as a secondary baseline rather than an experimental condition. This baseline is particularly useful for studying the effects of T4 and T5, since---unlike the main baseline T1---it utilizes a reviewing procedure based on pairwise comparisons.

T4 builds upon T3 by including the component of algorithmic assistance. Instead of asking respondents to review randomly selected pairs of apartments, the decision aid selects the pairs of apartments that the respondent will review. 

\begin{description}
    \item[T4:] While T3 presents respondents with random pairs of apartments, T4 selects \emph{pairs} where respondents' estimates are \emph{not aligned with the majority's view}. Specifically, we implicitly provided machine assistance by asking participants to review pairs of apartments for which their initial price estimates did not align with most people's comparative valuations of those apartments. We trained a model to predict the majority's comparative valuations of apartment prices using a dataset of human-annotated pairwise comparisons of apartments that we gathered, as described in Section \ref{subsec:decision_aids}.
\end{description}

Treatment T5 builds upon T4 by additionally providing explicit algorithmic advice. The addition of explicit advice makes the decision aid used in T5 more similar to the decision aids that are typically considered in the machine-assisted decision-making literature, such as the decision aid used in T2.

\begin{description}
    \item[T5:] The format of the review phase and the pair selection procedure remained the same as in T4, but we additionally explicitly informed people about the difference between their initial estimates and the predicted comparative valuations of most people.
\end{description}

\xhdr{Hypotheses}
We leverage prior research on machine-assisted decision-making in psychology and HCI (reviewed in Section \ref{sec:related_work}) to form hypotheses about the effects of our interventions T2, T4 and T5, compared to the control condition T1. As mentioned above, we treat T3 as a second baseline. We do not hypothesize about the effects of T3, and only study its effects exploratively.

Prior work has demonstrated that algorithmic decision aids can influence people's decisions \cite{grgic2019human}. In line with these findings, in H1 we hypothesize that our decision aids will prompt respondents to revise their estimates. Specifically, we hypothesize that compared to the control condition T1, our interventions T2, T4 and T5 will lead to:
\begin{description}
    \item[H1:] A higher number of decisions updated in the review phase.   
    \item[H1':] A higher propensity to update decisions for the particular apartments shown in the review phase.
\end{description}

We include both hypotheses H1 and H1' since they capture different aspects of the interventions' effects on people's propensity to update decisions. H1 focuses on the overall effect of the intervention across all apartments, while H1' captures the effectiveness in prompting respondents to review their estimate for specific apartments that are shown in the review phase. (These are the same for T1 and T2.) The effect captured by H1 may be deemed more important in settings where the goal is to maximize the overall effect across all apartments, while H1' may be more appropriate if we are interested in measuring engagement with the algorithmic assistance.

Furthermore, past research has demonstrated that accurate decision aids help increase the accuracy of people's decisions \cite{grgic2022taking, yin2019understanding}. In H2, we hypothesize that our decision aids---which, as discussed in Section \ref{subsec:decision_aids}, exhibit high predictive accuracy---will do the same. Namely, we hypothesize that compared to the control condition T1, our interventions T2, T4 and T5 will result in:
\begin{description}
    \item[H2:] A higher accuracy of post-review decisions.
\end{description}

Finally, we expect the decision aids to lead to an increase in the consistency between people's responses. We hypothesize that compared to T1, the interventions T2, T4 and T5 will lead to:
\begin{description}    
    \item[H3:] A higher degree of consistency between the post-review decisions of different respondents.
\end{description}

\xhdr{Dependent Variables}
For each apartment we measured the respondents' pre-review estimates and post-review estimates. Note that in treatments T3--T5, respondents were not given an opportunity to update their estimates for some apartments. In those cases, we defined their post-review estimate to be equal to their pre-review estimate.

To test our hypotheses, we formed dependent variables based on these measurements as follows:
\begin{description}
    \item[H1 magnitude:] Absolute difference between pre and post-review estimates.
    \item[H1' magnitude:] Absolute difference between pre and post-review estimates, limited to only those apartments shown during the review phase. 
    \item[H1 binary:] 0 if the pre and post-review estimates are equal, otherwise 1. 
    \item[H1' binary:] 0 if the pre and post-review estimates are equal, otherwise 1, limited to only those apartments shown during the review phase.
    \item[H2:] Difference between pre-review error and post-review error. The pre- and post-review errors are calculated as the absolute difference between the respondent's estimate and the ground truth for a given apartment. Intuitively, this measure captures the treatments' effects on the degree of agreement between respondents' estimates and the true apartment prices. 
    \item[H3:] Difference between pre-review inconsistency and post-review inconsistency. The pre- and post-review inconsistency are calculated as the absolute difference between the respondent's estimate and the average estimate (namely, the mean value of all respondents' estimates) for a given apartment. Intuitively, this measure captures the treatments' effects on the degree of agreement between the estimates of different respondents.
\end{description}

\xhdr{Analysis}
In Section \ref{sec:results}, we report the findings of our confirmatory analyses related to hypotheses H1--H3. In the text, we report the findings of our statistical hypothesis testing, accompanied with plots that illustrate our findings using descriptive statistics. 

To test our hypotheses we rely on linear mixed models. We note that in hypotheses H1 binary and H1' binary our dependent variables are binary. The choice between using a linear and a logistic regression in such settings has been much debated in prior work, and we refer the readers to the work of \mbox{\citet{hellevik2009linear}} for an in-depth discussion on this topic. Hence, we replicated our analyses for binary dependent variables using both a linear and a logistic regression. The results are qualitatively the same for both models. In the paper, we report the results of the linear regression for ease of interpretation of the coefficients and consistency with other hypotheses, in line with our pre-registration.

Due to our repeated measures design, we include crossed random effects to account for differences between participants and apartments. We compared models that include a participant or apartment random effects term to nested models that do not include these random effects terms. To do so, we used likelihood-ratio tests. The likelihood-ratio tests confirmed that including participant and apartment random effects terms makes a difference---i.e., there is a significant amount of variation between participants and apartments accounted for by the random intercepts (p-val <0.001, for all 6 models, and both for participant and apartment random effects terms). To alleviate convergence issues of models with crossed random effects, we initialize the starting values of the parameters to the estimated parameters of a simpler model---a linear mixed model with a random effects term for participants only. We further note that we compared our models to fixed effects models with two-way clustering of standard errors with respect to apartments and participants, trained utilizing the \texttt{reghdfe} Stata package \mbox{\cite{correia2017reghdfe}} that implements the estimator described in \mbox{\citet{Correia2017:HDFE}}. We found the results of both approaches to lead to consistent findings across all hypotheses.

The dependent variables vary across hypotheses as described in the previous subsection. In all of the models, the experimental conditions are used as the independent variables. We one-hot encode the five experimental conditions T1--T5 using four binary variables corresponding to T2--T5, and we treat T1 as the reference category. That is, our models' unstandardized regression coefficients capture how the effects of treatments T2--T5 differ from the effects of T1. To compare the effects of other pairs of treatments we utilize Wald tests to test the equality of the corresponding coefficients. For the Wald tests, we report Bonferroni-adjusted p-values to account for the multiple comparisons problem.

In the appendix we conduct an additional exploratory analysis of our data. There we report a series of descriptive statistics, including results related to the effect of the treatments on other measures of accuracy and consistency.

\begin{figure}[t]
    \centering
    \includegraphics[width=0.65\textwidth,trim={7.5cm 9.5cm 7.5cm 3cm},clip]{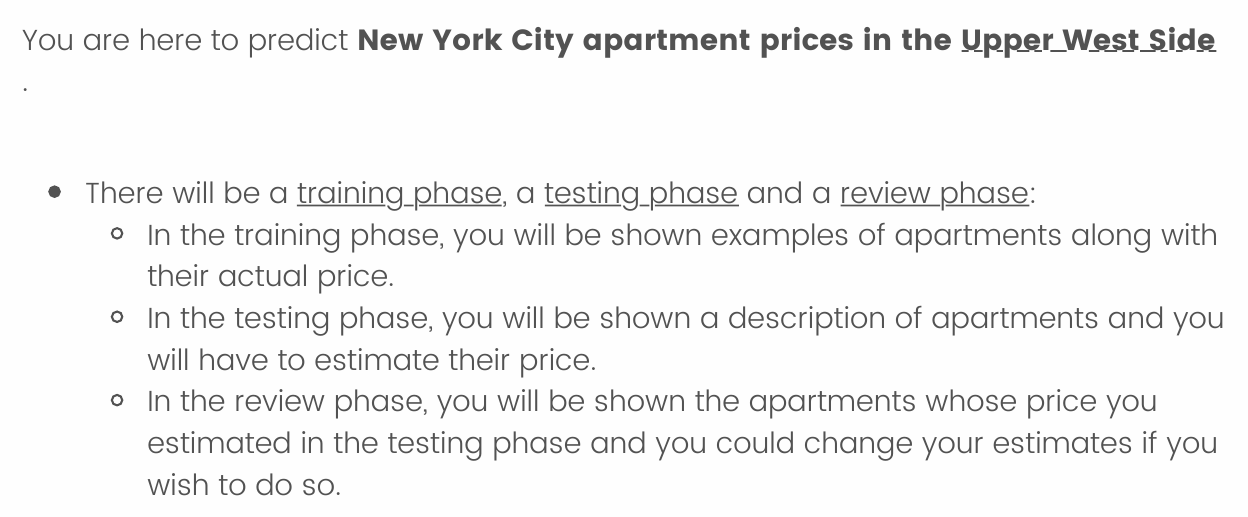}
    \caption{Description of the experimental design shown to participants at the beginning of the experiment.}
    \label{fig:intro_text}
\end{figure}

\begin{figure}[t]
    \begin{subfigure}[t]{0.49\textwidth}
         \centering
         \includegraphics[width=\textwidth,trim={7.5cm 4.25cm 8.5cm 3.25cm},clip]{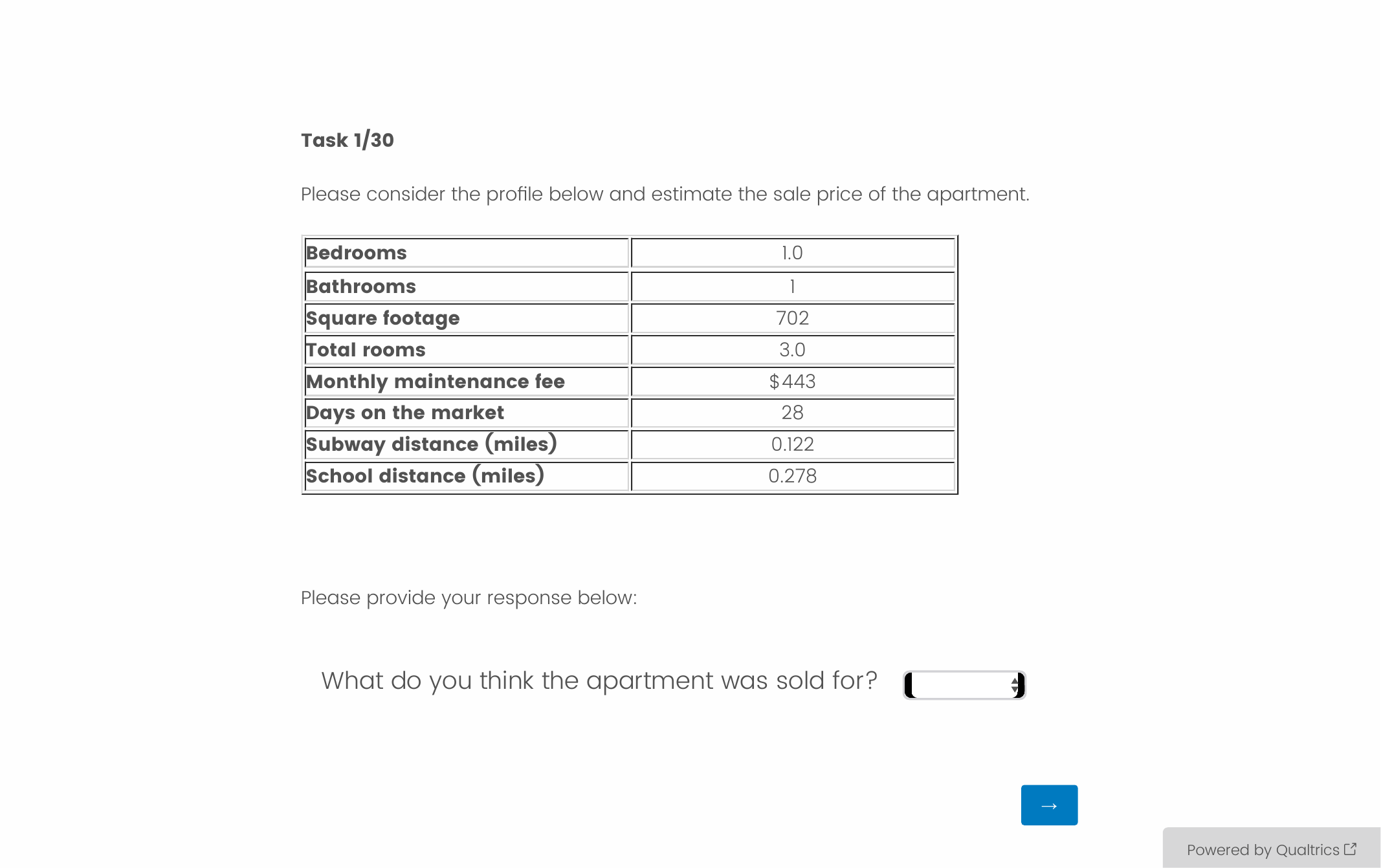}
         \caption{Survey instrument used in the tutorial and for gathering respondents' pre-review estimates, across all treatments.}
         \label{fig:pre_review_qs}
    \end{subfigure}
    \hfill
    \begin{subfigure}[t]{0.49\textwidth}
         \centering
         \includegraphics[width=\textwidth,trim={5cm 5cm 11cm 1cm},clip]{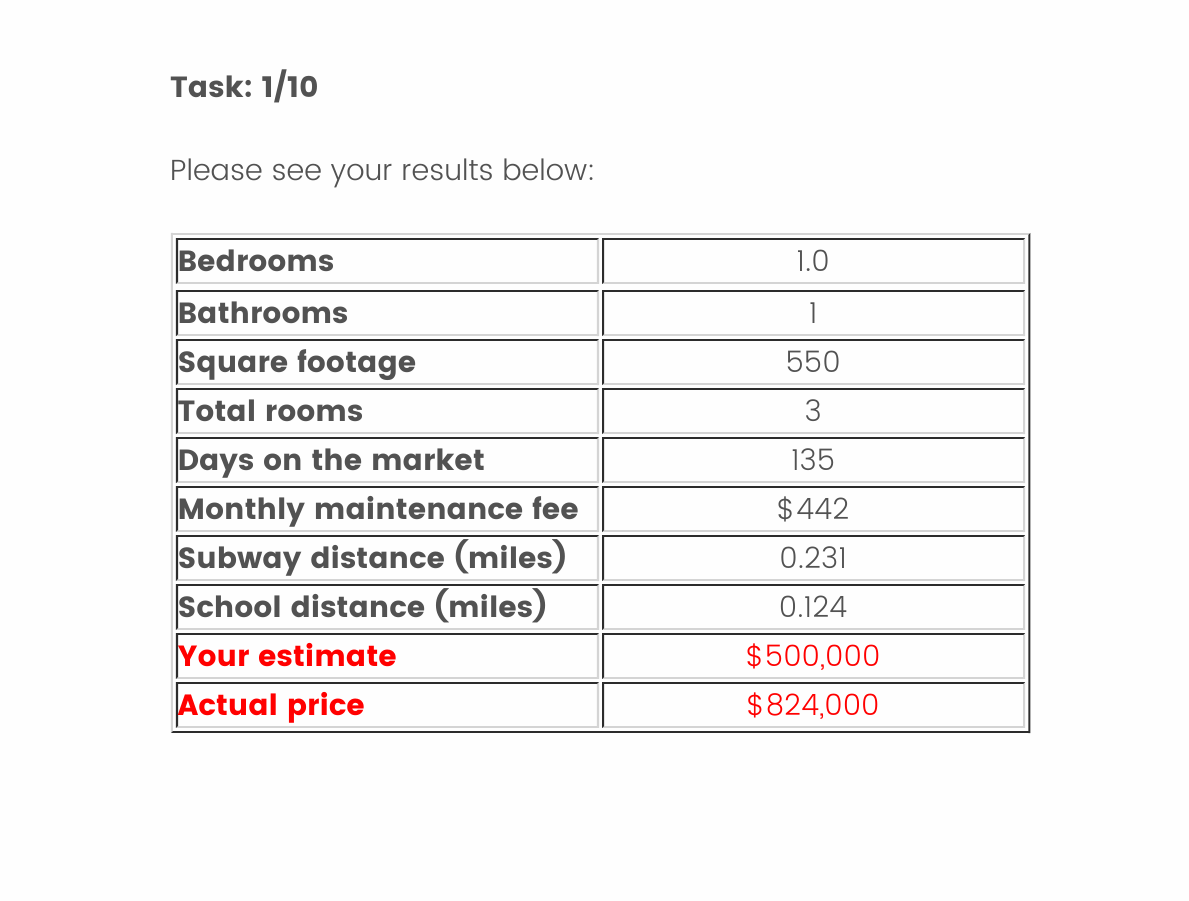}
         \caption{Feedback provided to respondents during the tutorial, in all experimental conditions.}
         \label{fig:tutorial_feedback}
    \end{subfigure}
    \hfill
    \begin{subfigure}[t]{0.49\textwidth}
         \centering
         \includegraphics[width=\textwidth,trim={7.5cm 2.25cm 8.5cm 0cm},clip]{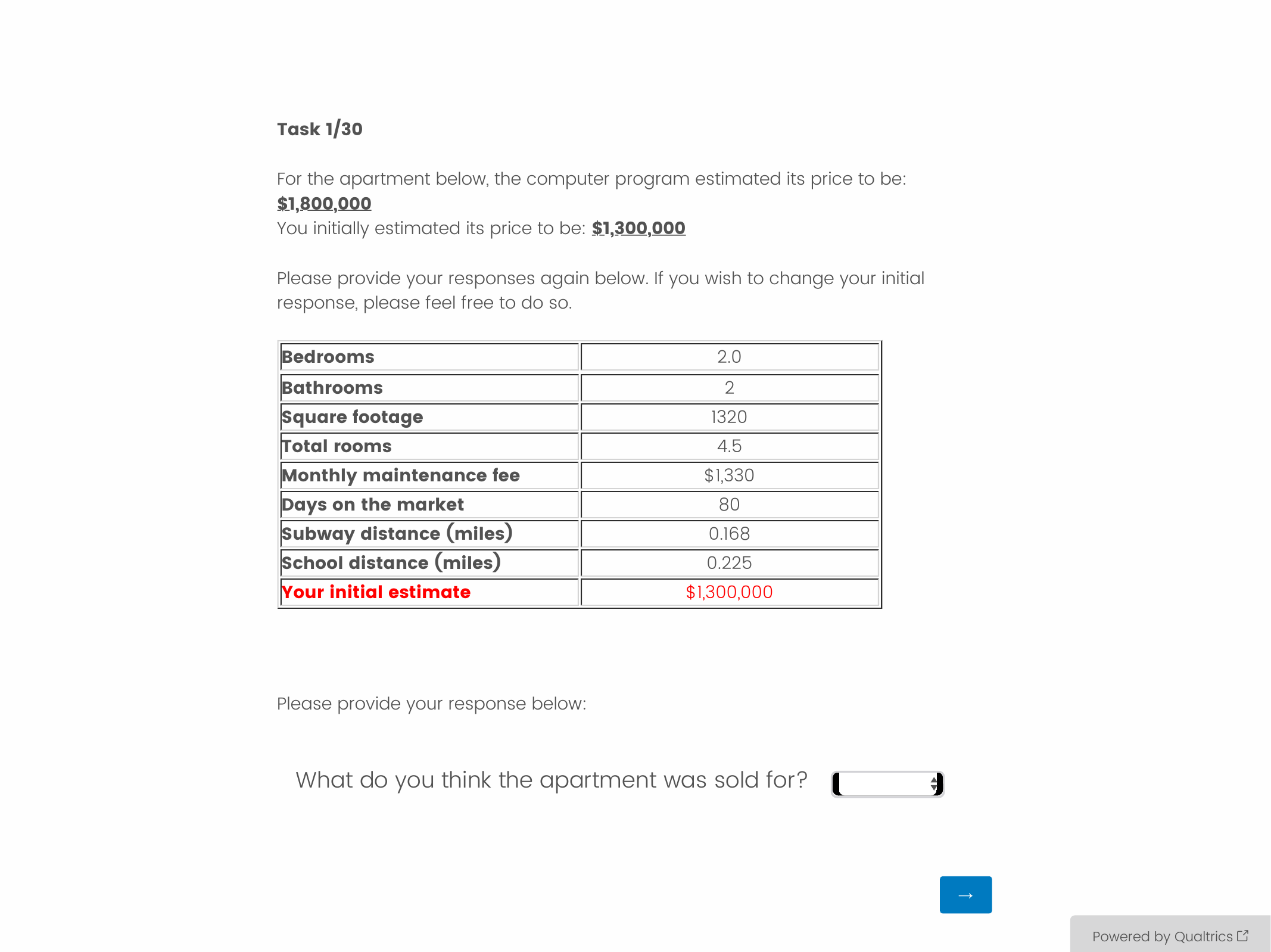}
         \caption{Survey instrument used for gathering participants' post-review estimates in T2. In T1, the machine prediction was omitted.}
         \label{fig:treatments_absolute}
    \end{subfigure}
    \hfill
    \begin{subfigure}[t]{0.49\textwidth}
         \centering
         \includegraphics[width=\textwidth,trim={11cm 3.25cm 11.5cm 1.5cm},clip]{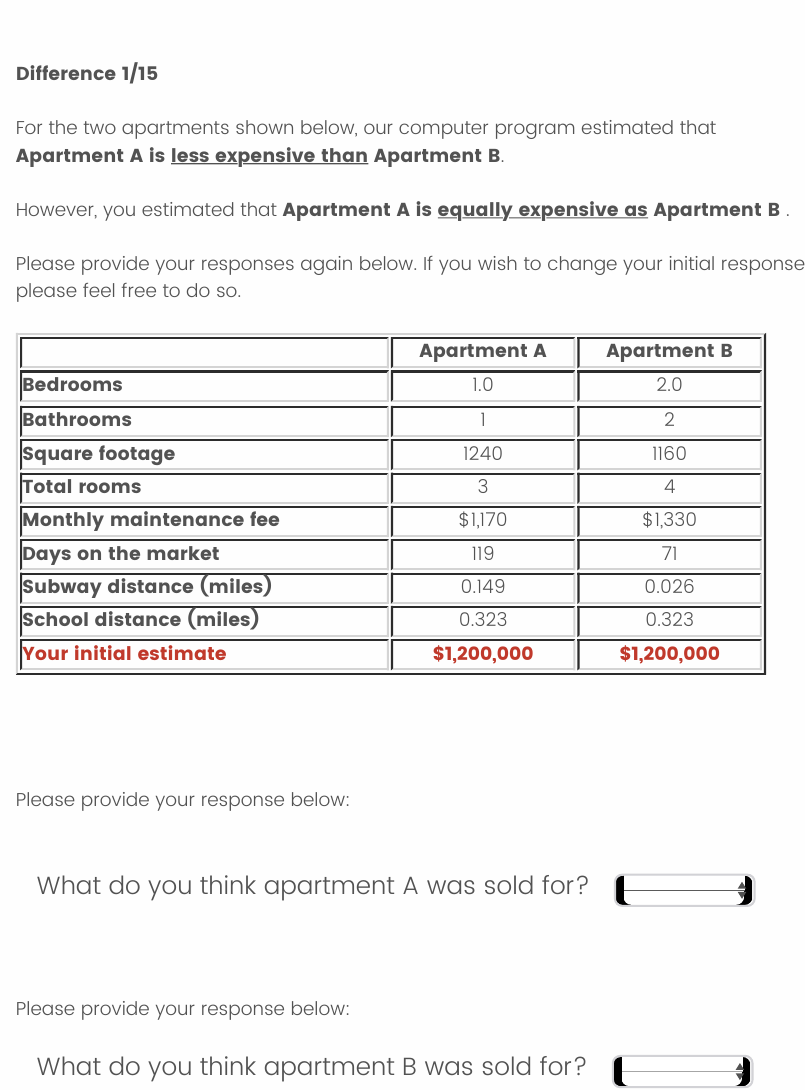}
         \caption{Survey instrument used for gathering participants' post-review estimates in T5. In T3 and T4, the machine prediction was omitted.}
         \label{fig:treatments_pairwise}
    \end{subfigure}
    \hfill
    \caption{Stimulus material.}
    \label{fig:stimulus_material}
\end{figure}

\subsection{Stimulus Material} \label{subsec:stimulus_material}
Upon opening the study link through the Prolific interface, participants were randomly assigned to one of the five experimental conditions. All participants first completed an online consent form and entered their Prolific worker ID. Next, participants were shown an introductory text describing the task (Figure \ref{fig:intro_text}). 

Following the approach of \citet{poursabzi2021manipulating}, participants were asked to complete a tutorial in order to familiarize themselves with real estate prices in New York City. The tutorial consisted of the same ten apartments that were utilized by \citet{poursabzi2021manipulating}. The ten apartments were shown in random order, and for each apartment respondents were first asked to estimate its price based on its brief description (Figure \ref{fig:pre_review_qs}), and were then informed about the apartment's actual listing price (Figure \ref{fig:tutorial_feedback}). 

Next, we gathered the first part of our experimental data---the respondents' pre-review price estimates. We asked all participants to estimate the prices of the same 30 apartments. The 30 apartments were selected uniformly at random from the 387 apartments in the dataset, excluding the 10 apartments utilized in the tutorial. The set of apartments was kept constant across all experimental conditions and respondents. Throughout the experiment, in all five treatments, the apartments were shown in random order to avoid order bias \cite{redmiles2017summary, groves2011survey}. The phrasing and format of the questions and response options were identical to the tutorial (Figure \ref{fig:pre_review_qs}), except that we did not provide respondents with information about the apartments' true listing price after they reported their estimates.

The respondents were then asked to respond to one simple instructed response item, which served as an attention-check question. Specifically, respondents were asked to ``Please respond to this question by selecting Somewhat disagree as the answer,'' using a 5-point Likert scale as the response options. Similar instructed response items are commonly used for quality assurance purposes in online surveys, as a means of identifying inattentive or careless respondents \cite{meade2012identifying}.

Next, we gathered the second part of our experimental data---the respondents' post-review price estimates. Respondents were presented with a text describing the experimental condition they were assigned to, i.e., they were informed about the procedure they will follow in the review phase. Participants were asked to review their initial responses in one of five different ways, based on the experimental condition they were randomly assigned to. The experimental conditions T1--T5 are described in Section \ref{subsec:experimental_design} ``Experimental Conditions'' and depicted in Figures \ref{fig:treatments_absolute} and \ref{fig:treatments_pairwise}. All respondents reviewed 30 of their estimates. However, depending on the treatment, respondents reviewed all of their initial estimates one by one (T1 and T2) or a subset of their initial estimates in a series of pairwise comparisons with possible repetition of the apartments, up to 3 times, in multiple pairs (T3, T4 and T5).

Finally, we gathered participants' feedback about their experience of participating in the experiment. Namely, we asked respondents to tell us how much they agree with the following statements on a 5-point Likert scale from ``Strongly agree'' to ``Strongly disagree'': (i) The study was interesting, (ii) I would like to take part in a similar study in the future, (iii) The questions were easy to understand, (iv) The study was too long. At the end of the study the respondents also had the option to provide additional comments that they wanted to share with the researchers.

\begin{table}[t]
\centering
\begin{tabulary}{\linewidth}{LRR}
    \toprule
    {\bf Demographic Attribute} & {\bf Sample} & {\bf Census}\\
    \hline
    <35 years & 56.4\% & 46\%\\
    35--54 years & 32.6\% & 26\%\\
    55+ years &  10.8\% & 28\%\\
    \hline
    Male & 50\% & 49\%\\
    \hline
    Asian & 10.3\% & 6\% \\
    Black & 10.5\% & 12\% \\
    Hispanic & - & 18\% \\
    Mixed & 6.1\% & - \\
    White & 68.5\% & 61\% \\
    Other & 4.6\% & 4\% \\
    \bottomrule
\end{tabulary}
\caption{Demographics of our study sample, compared to the 2019 U.S. Census~\cite{census_acs}.}
\label{tab:demographics}
\end{table}

\begin{figure}[t]
    \centering
    \includegraphics[width=0.55\textwidth]{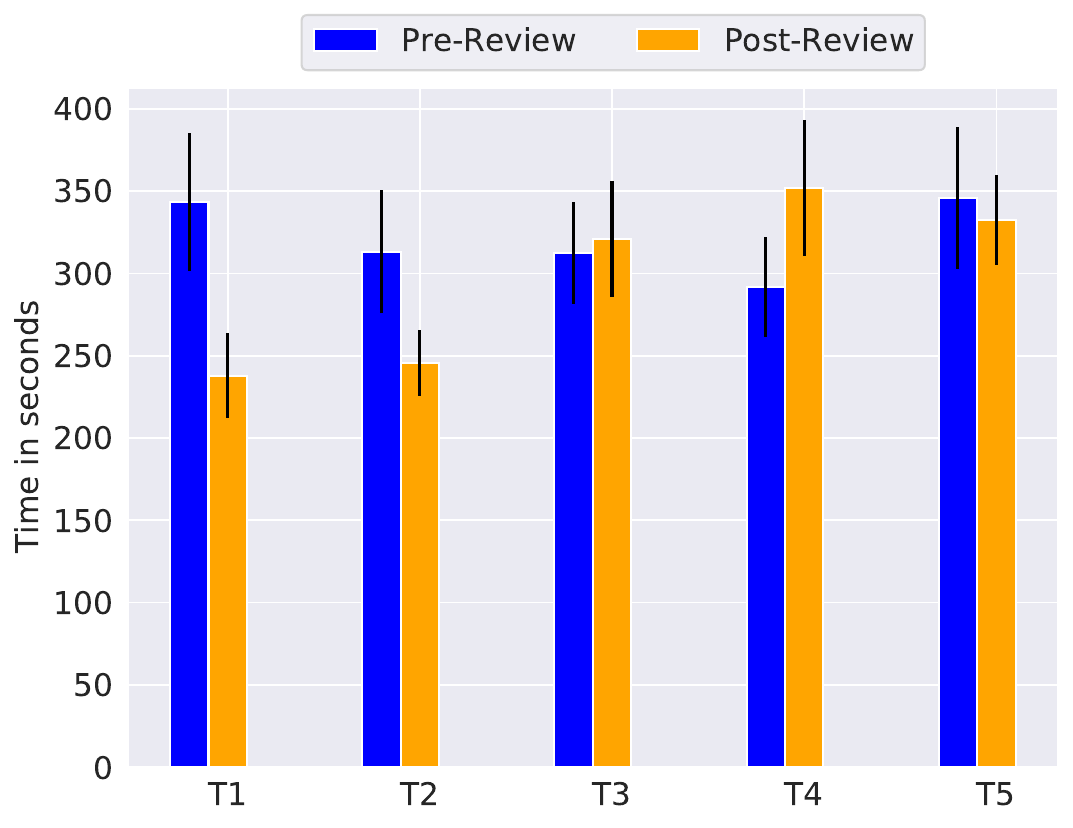}
    \caption{Average duration of the experiment, per experimental condition, and per experimental phase. The experimental conditions T1--T5 are shown on the x-axis. The values for the pre-review experimental phase are shown in blue, while the post-review values are shown in orange. We report mean values calculated across respondents $\pm$ 1.96 standard errors of the mean (SEM).}
    \label{fig:study_duration}
\end{figure}

\subsection{Data Collection} \label{subsec:data_collection}
We recruited participants from Prolific---an online crowdsourcing platform which caters to scientific researchers \cite{palan2018prolific}. Using Prolific's built-in pre-screening capabilities, we targeted respondents who: (i) are located in the US, (ii) have participated in at least 10 Prolific studies in the past, and (iii) have an approval rate of at least 95\% on these past studies. We additionally utilized Prolific's option to provide a sample of respondents that is balanced with respect to gender, due to the current gender imbalance on the platform \cite{prolific2021gender}. 

Our goal was to recruit sufficiently many participants to detect medium-sized effects (Cohen's $d = 0.5$) at the significance level of $\alpha = 0.05$ with power $\beta = 0.95$. Using the statistical software G*Power \cite{faul2007g,faul2009statistical}, we calculated that a conservative Wilcoxon-Mann-Whitney two-tailed test requires 110 respondents per treatment group to detect effects of the size, significance level and power of interest. In our study, we have five experimental conditions, leading us to a minimum sample size of 550 respondents. To account for possible exclusions, incomplete or missing responses, we increased this estimate by 20\% to 660 respondents. In Appendix \ref{subsec:appendix_details}, we report the values of Cohen's d calculated on the gathered dataset.

We recruited a total of 660 participants from Prolific, over the course of several days (30th November -- 3rd December 2022) in order to minimize sampling bias that could occur due to the day in the week or the time of day \cite{casey2017intertemporal}. Participants were paid GBP 3.1 for taking part in the study. On average, participants were paid GBP 12.03 per hour, i.e., approximately USD \$14.80 per hour---well above the federal minimum wage of USD \$7.25.  The median study completion time was 15 minutes and 28 seconds. The duration of the experiment varied across treatments, as depicted in Figure \ref{fig:study_duration}. As expected, there were no statistically significant differences between the average time taken to complete the pre-review phase of different experimental conditions. However, we observe significant differences across treatments in the review phase. Specifically, the review phase in T1 and T2---where respondents reviewed decisions one-by-one---took significantly less time than in T3--T5, where respondents reviewed pairs of decisions. When comparing the duration of the pre-review and review phase, T1 and T2 led to a significant increase in speed. On the other hand, the review phase in T4, where meaningfully selected pairs were presented without explicit advice, took more time compared to the pre-review phase. In T3 and T5 both the pre-review phase and the review phase took a similar amount of time to complete.

We report the demographics of our sample in Table \ref{tab:demographics}. Since none of our hypotheses rely on demographic data, we did not ask our respondents to complete a demographics survey, in order to minimize the duration of our experiment, and to align with the data minimization principle. Hence, we report the data about our participants that we had access to through the crowdsourcing platform Prolific. Please note that this demographic data was self reported by Prolific crowdworkers directly to Prolific. Compared to the US census, our sample is younger, in line with typical samples recruited via online crowdsourcing platforms \cite{huff2015these, ross2010crowdworkers, paolacci2010running}. In line with the gender balancing pre-screening criteria employed during sampling, our sample is balanced with respect to gender. In terms of ethnicity, we are not able to directly compare our sample to the US census, since Prolific's simplified ethnicity prompt did not offer "Hispanic" as a response option. However, we note that Asian respondents are slightly over-represented and Black respondents are slightly underrepresented compared to the US census data.

Upon completing the study, participants were asked to provide feedback about their experience of taking part in this study. Most was positive. On a 5-point Likert scale from ``Strongly agree'' (coded as 5) to ``Strongly disagree'' (coded as 1), participants agreed with the statements 
``The study was interesting'' ($\mu = 4.1\pm 1.0$ ), 
``I would like to take part in a similar study in the future'' ($\mu = 4.4 \pm 0.9$), and
``The questions were easy to understand'' ($\mu = 4.6 \pm 0.8$), 
while they neither agreed nor disagreed with the statement 
``The study was too long'' ($\mu = 2.7\pm 1.1$). 

For the purposes of our analyses, we excluded all responses from participants who did not complete the full study (i.e., missing or incomplete responses), or who failed the instructed response attention check questions. A total of 17 respondents (2.6\%) failed the attention check, leaving us with a final sample of 643 respondents.

\subsection{Decision Aids} \label{subsec:decision_aids}

\subsubsection{Developing the Decision Aid Utilized in T2}\hfill\\
For T2, we developed a typical example of a decision aid within the judge-advisor system (JAS) paradigm \cite{bonaccio2006advice}. It is trained to accurately estimate real estate prices. The algorithm's accurate advice can then help steer people towards making accurate estimates of real estate prices. That is, this decision aid was designed to be aligned with our hypothesis H2. 

While it was not explicitly designed to increase people's consistency (H3), we still expect that it will succeed in doing so. If the decision aid successfully steers people towards its advice, it will trivially lead to an increase in inter-respondent consistency. 

\xhdr{Training Procedure}
We trained a linear regression model that used the apartment's attributes as independent variables (full list of features shown in Figure~\ref{fig:pre_review_qs}), and the apartment price as the dependent variable. We normalized the independent variables to have zero mean and unit variance. We considered models with L1 (lasso) and L2 (ridge) regularization and without regularizers, and picked the regularization hyperparameter values which resulted in the highest coefficient of determination ($R^2$) on a 20\%  held out validation set. 

\xhdr{Evaluation}
The model without any regularizer yielded the highest $R^2$ value: 85.86 on a test set comprised of 30\% of the data. The mean absolute error of this model was \$143,200. Amongst all of the features used in the final model, ``Square footage'' exhibited the strongest positive correlation with apartment price (with a weight of 1.08), while ``Total rooms'' exhibited the strongest negative correlation with price (with a weight of -0.3). 

\xhdr{Format of Machine Advice}
In T2, we provided this model's estimates rounded to the nearest $\$$100,000 as machine advice, for ease of interpretation.

\subsubsection{Developing the Decision Aid Utilized in T4 and T5}\hfill\\
The decision aid developed for T4 and T5 differs from the decision aids typically studied in the machine-assisted decision-making literature. Instead of predicting apartments' true prices, it is trained to predict people's comparative valuations of apartments. The algorithm's advice can then help steer people towards making estimates that are consistent with other people's estimates. That is, this decision aid was designed to be aligned with our hypothesis H3.

Despite being trained only to predict human comparative valuations of apartments, the wisdom of the crowd enabled this decision aid to accurately predict the true comparative valuations of apartments as well. Therefore, we expect it to also succeed in increasing the accuracy of people's responses (H2).

\xhdr{Data Gathering}
In order to build this tool, we gathered a dataset of human comparative valuations of apartments. We randomly selected 1000 unique pairs of apartments from our dataset and split them into 40 batches of 25 pairs. We recruited a total of 850 Prolific workers, who were randomly assigned to one of the batches. After excluding respondents who failed the attention-check question, we were left with 806 participants. From these, we excluded the last 6 responses so that each batch was labelled by exactly 20 participants. We gathered the data over several days (8th November -- 11th November 2022) to minimize any bias caused by the time at which the data was gathered \cite{casey2017intertemporal}. The participants were paid GBP 2 for taking part in the study, resulting in an average hourly rate of approximately USD \$13.50. 

The stimulus material and the experimental procedure were similar to the ones used in the main experiment, described in \ref{subsec:stimulus_material}. The participants completed a consent form and entered their Prolific worker IDs, prior to observing an introductory text similar to the one shown in Figure \ref{fig:intro_text}. The participants then completed the same tutorial as in the main experiment, in which they were asked to estimate the prices of 10 apartments prior to observing their actual listing price, as shown in Figure~\ref{fig:tutorial_feedback}. Finally, respondents were asked to compare pairs of apartments, which were presented as shown in Figure~\ref{fig:treatments_pairwise}. Specifically, they were asked to estimate if ``Apartment A'' or ``Apartment B'' were more expensive. Additionally, they were asked how confident they were in their estimate on a 5 point Likert scale, ranging from ``Completely guessing'' to ``Completely confident.'' To avoid order bias \cite{redmiles2017summary, groves2011survey}, both the order of the 30 pairs of apartments and the order of apartments within a given pair were randomized.

\xhdr{Training Procedure}
Using this data, we trained a cross validated logistic regression classifier with L2 regularization to predict which apartment is perceived as more expensive in a given pair. To form the independent variables for our classifier, we subtracted the features of pairs of apartments (shown in Figure~\ref{fig:pre_review_qs}) from one another. We then normalized them to have zero mean and unit variance. As the dependent variable we used the confidence weighted majority votes of the participant's responses. E.g., if a participant was ``Completely guessing'' their vote would count as $\frac{1}{5}$ and if they were ``Completely confident'' it would count as 1. 

\xhdr{Evaluation}
In the trained classifier ``Square footage'' was the most important feature, i.e., it had the largest absolute weight (36.18). The second most important feature was ``Maintenance cost'' (18.3). Our classifier predicted people’s comparative valuations of apartments with an accuracy of 98.4\%, cross-validated on five randomly chosen 30\% test sets.

While this classifier was trained to predict \emph{people's} comparative valuations of apartments, it also exhibited a high degree of accuracy in predicting apartments' \emph{true} comparative valuations. Namely, the accuracy of predicting the pairwise order of apartments with respect to the ground truth prices was 88.5\%. In practice, this tool would have an even higher accuracy since we prioritized giving advice for pairs where the tool had high confidence. Such pairs demonstrated a higher accuracy compared to those pairs where the decision aid had low confidence. The high accuracy of this tool led us to hypothesize that the decision aid used in Treatments T4 and T5 can increase not only the consistency (H3) of people's responses, but also the accuracy (H2) of the participant's estimates, despite being \emph{trained only on human annotations instead of ground truth labels}.

\xhdr{Format of Machine Advice}
In order to provide assistance to participants in T4 and T5, we used the aforementioned classifier. We utilized this decision aid to identify pairs of apartments for which people's estimates did not align with the majority's comparative valuation. We prioritized giving advice for pairs where (i) our classifier was able to accurately predict a typical person's comparative valuation, and (ii) the predicted ordering did not match the respondent's ordering.

First, we converted the 30 apartments used in the main study into 435 pairs by taking all possible combinations, and predicted which of the apartments in each pair would be perceived as more expensive by most people. Then we ordered the pairs in a decreasing order with respect to the classifier's confidence. In T4 and T5 we iterated through this list, and asked participants to review their initial decisions which did not align with our classifier's predictions. E.g., if they initially estimated that Apartment A cost $\$600,000$ and that Apartment B cost $\$900,000$, while the classifier predicted that most people would perceive Apartment A as more expensive than Apartment B, participants could have been asked to review this pair of apartments. Participants were asked to review 15 pairs of apartments from this list. A single apartment was limited to appear in at most 3 pairs, to avoid negative reactions from repeatedly being presented with the same information. While selecting the pairs to show respondents, we took into account the decisions they may have updated during the review process. 
\begin{figure}[t]
    \begin{subfigure}[t]{0.47\textwidth}
         \centering
         \includegraphics[width=\textwidth]{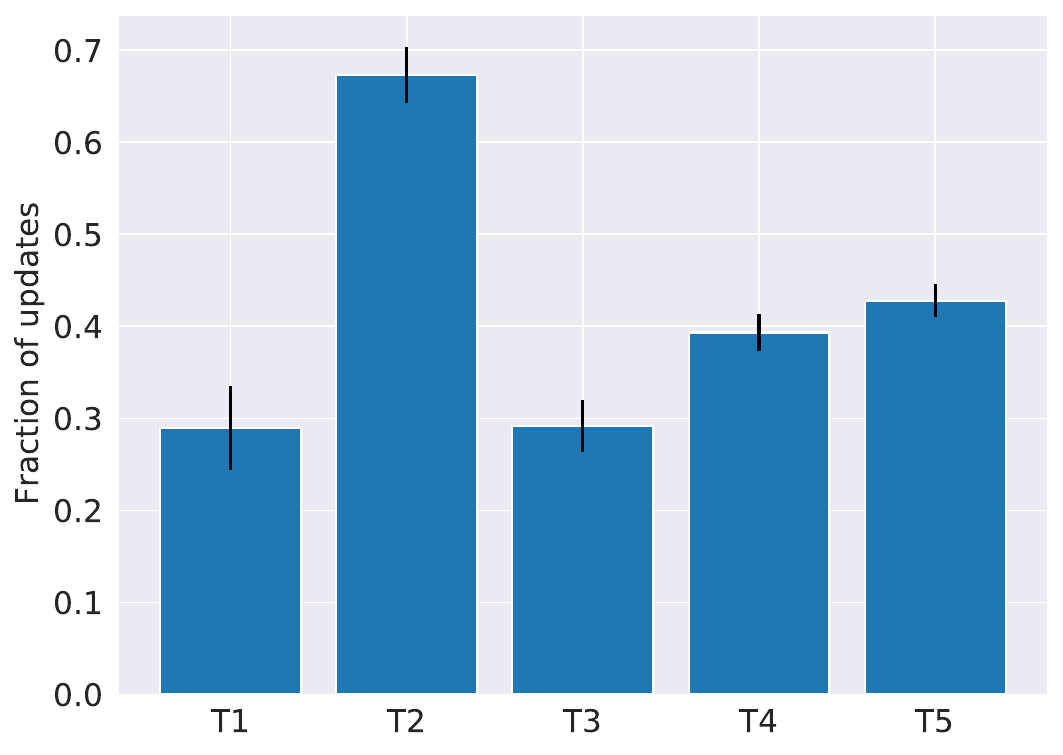}
         \caption{H1 binary: Fraction of decisions updated. The y-axis shows the fraction of the 30 initial decisions that were updated in the review phase.}
         \label{fig:H1_binary}
    \end{subfigure}
    \hfill
    \begin{subfigure}[t]{0.50\textwidth}
         \centering
         \includegraphics[width=\textwidth]{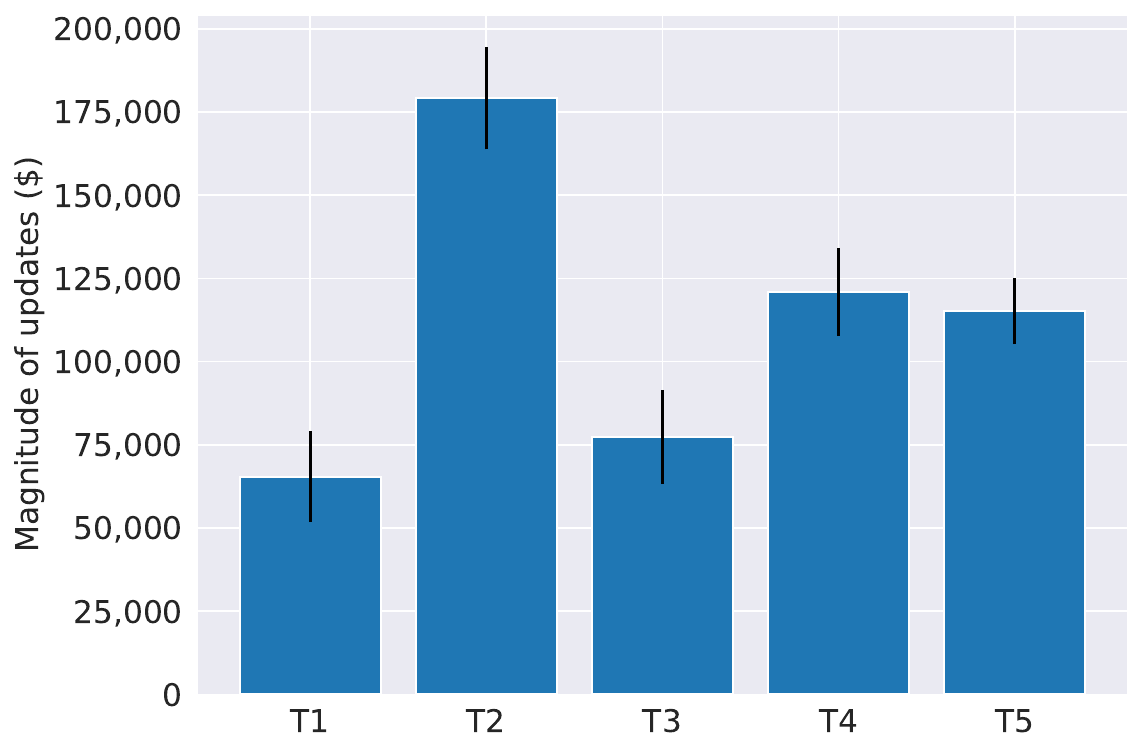}
         \caption{H1 magnitude: Magnitude of updates. The y-axis shows by how much the 30 initial decisions were updated in the review phase.}
         \label{fig:H1_magnitude}
    \end{subfigure}
    \hfill
    \caption{H1: Effect of the interventions on people's propensity to update decisions, across all 30 apartments. The experimental conditions T1--T5 are shown on the x-axis. We report mean values calculated across respondents $\pm$ 1.96 standard errors of the mean (SEM).}
    \label{fig:H1}
\end{figure}

\begin{table}[t]
\centering
\def\sym#1{\ifmmode^\textbf{#1}\else\(^\textbf{#1}\)\fi}
\begin{tabular}{l*{6}{c}}
    \toprule
        & \textbf{H1}: change, & \textbf{H1'}: change, & \textbf{H1}: change, & \textbf{H1'}: change, & \textbf{H2}: & \textbf{H3}: \\
    & bin., overall & bin., specific & mag., overall & mag., specific & accuracy & consistency \\
    \hline \\
    [-0.75em]
    \textbf{T2} & 0.383\sym{***} & 0.383\sym{***} & 113634.1\sym{***} & 113634.1\sym{***} & 97299.5\sym{***} & 119916.8\sym{***}\\
    & (0.0216) & (0.0260) & (9608.2) & (13339.4) & (6022.1) & (6840.6) \\
    [0.5em]
    \textbf{T3} & 0.00179 & 0.157\sym{***} & 11897.4 & 52257.4\sym{***} & -7141.5 & -10044.1 \\
    & (0.0215) & (0.0264) & (9589.6) & (13489.2) & (6010.5) & (6827.3) \\
    [0.5em]
    \textbf{T4} & 0.103\sym{***} & 0.395\sym{***} & 55471.8\sym{***} & 140695.3\sym{***} & 25568.7\sym{***} & 24214.8\sym{***}\\
    & (0.0218) & (0.0269) & (9685.0) & (13713.6) & (6070.3) & (6895.3) \\
    [0.5em]
    \textbf{T5} & 0.138\sym{***} & 0.449\sym{***} & 49783.9\sym{***} & 130046.7\sym{***} & 20490.2\sym{***} & 24042.4\sym{***}\\
    & (0.0216) & (0.0266) & (9608.2) & (13596.1) & (6022.1) & (6840.6) \\
    [0.5em]
    \textbf{Cons.} & 0.290\sym{***} & 0.290\sym{***} & 65461.5\sym{***} & 65461.5\sym{***} & 7563.6 & 2271.4 \\
    & (0.0212) & (0.0216) & (11207.0) & (13277.9) & (6679.6) & (5555.7) \\
    [0.5em]
    \hline
    \(N\) & 19290 & 14659 & 19290 & 14659 & 19290 & 19290 \\
    \bottomrule
\end{tabular}
\caption{Linear mixed models with crossed random effects for participants and apartments. The dependent variables for different hypotheses are described in Section \ref{subsec:experimental_design}. In all six models, the four independent variables T2--T5 correspond to a one-hot encoding of the experimental conditions, and T1 is treated as the reference category. I.e., intuitively, the row ``Cons.'' shows the estimated value of the constant term (or intercept) that corresponds to the effects of treatment T1, while the rows T2--T5 show how the effects of these treatments differ compared to T1. Hence, to reason about the effects of T2--T5, one needs to sum up the values of the constant term and the treatment of interest. \emph{N} denotes the number of data points used to fit a specific model. Each of our 643 respondents answered questions about 30 apartments, resulting in a total of 19290 data points. Please note that in the analysis for H1' some of the data points are discarded, as described in Sections \ref{subsec:experimental_design} and \ref{subsec:resultsH1'}. Standard errors are shown in parentheses. Statistical significance of coefficients is indicated as follows: * p < 0.05, ** p < 0.01, *** p < 0.001.}
\label{tab:regression}
\end{table}

\section{Results} \label{sec:results}
In this section, we present the results of our analysis. We compare the baseline reviewing procedure T1 to our interventions T2, T4 and T5, in terms of their effect on people's propensity to \emph{update} their initial estimates (H1 and H1'), and the \emph{accuracy} (H2) and \emph{consistency} (H3) of people's estimates.

\subsection{H1: Overall Change in Decisions} \label{subsec:resultsH1}
In all five experimental conditions, we observe that people update some of their 30 initial decisions in the review phase. However, the number of decisions that are updated and the magnitude of these updates varies substantially between the experimental conditions. Compared to the control condition T1 our interventions T2, T4 and T5 lead to a higher propensity to update decisions in the review phase. That is, \textbf{our results support H1}. This holds both in terms of the number of decisions that were updated (H1 binary) and the magnitude of the change (H1 magnitude).

\xhdr{H1 binary: Number of Decisions Updated}
Descriptively, we find that the fraction of decisions that are updated varies between treatments (first column of Table \ref{tab:regression} and Figure \ref{fig:H1_binary}). In both T1 and T3 people update approximately $29\%$ of their decisions, i.e., they update the estimated prices of 8.7 out of 30 apartments on average. In T4 and T5 people update a larger fraction of their decisions than in T1 and T3---close to $39\%$ (11.8/30 apartments) and $43\%$ (12.8/30 apartments) respectively. The treatment T2 has proven to have the strongest effect on people's propensity to update their decisions, with approximately $67\%$ of decisions (20.2/30 apartments) being updated.

These descriptive observations are corroborated by our statistical analyses. The regression in the first column of Table \ref{tab:regression} shows that all five treatments significantly influence human decisions. T2, T4 and T5 have a significantly stronger effect than T1, while the effect of T3 was not significantly different than that of T1. Subsequent Wald tests performed on the estimated model confirmed that T4 and T5 also have a stronger effect than T3 ($p<0.001$), but did not identify a significant difference between the effects of T4 and T5 ($p=0.35$). Finally, T2 was shown to have a significantly stronger effect than all of the other treatments ($p<0.001$). 

That is, we find that people are more likely to update their decisions when reviewing meaningfully selected pairs of apartments and when machine advice is provided.

\xhdr{H1 magnitude: Magnitude of Updates}
Our findings related to the magnitude of the changes are aligned with the findings about the number of decisions updated (third column of Table \ref{tab:regression} and Figure \ref{fig:H1_magnitude}). In T1, people update their decisions by approximately \$65,461 on average. In T3, the average update is close to \$77,359. The effect of both T1 and T3 is significantly different than zero ($p<0.001$), and the difference between these two treatments is not statistically significant ($p=0.214$). In T4, the average magnitude of the change was close to \$120,933. This is a significant increase compared to both T1 and T3 ($p<0.001$). In T5, people updated their decisions by \$115,245 on average, which is significantly more than T1 and T3 ($p<0.001$), but not significantly different than T4 ($p=1$). Finally, people changed their decisions by close to \$179,096 in T2. The magnitude of this change is significantly larger than in any of the remaining treatments ($p<0.001$). 

In short, we find that people update their decisions by a larger amount when they review them as a series of meaningfully chosen pairwise comparisons and when they observe machine advice.

\begin{figure}[t]
    \begin{subfigure}[t]{0.47\textwidth}
         \centering
         \includegraphics[width=\textwidth]{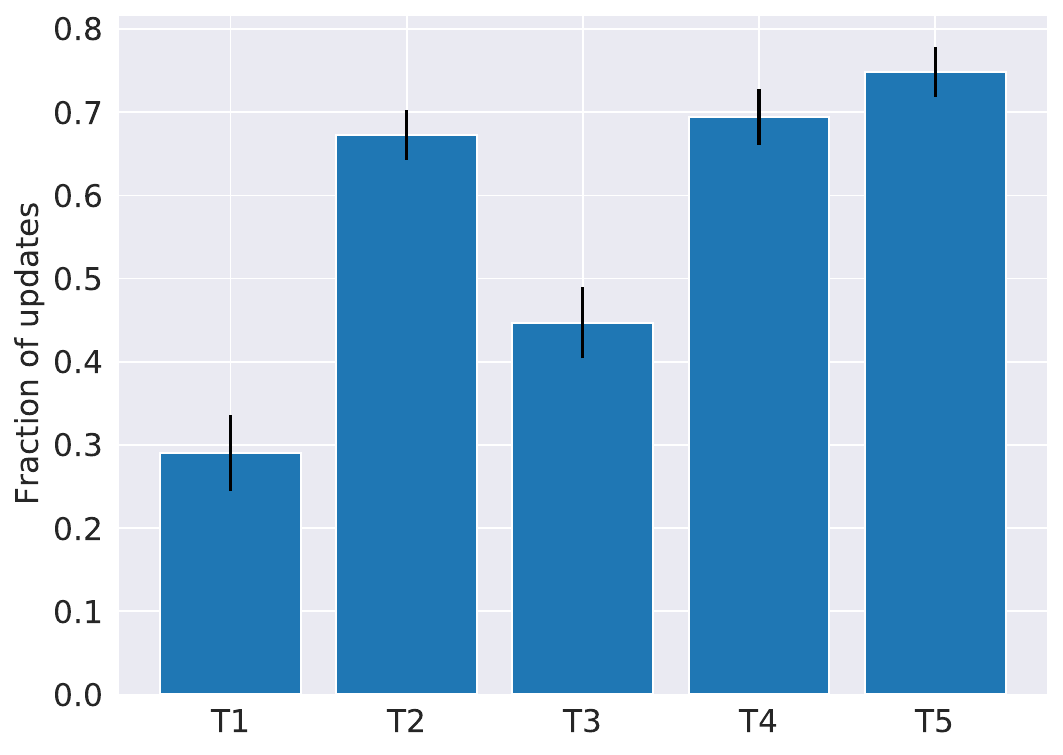}
         \caption{H1' binary: Fraction of decisions updated. The y-axis shows the fraction of the decisions shown in the review phase that were updated.}
         \label{fig:H1_prime_binary}
    \end{subfigure}
    \hfill
    \begin{subfigure}[t]{0.50\textwidth}
         \centering
         \includegraphics[width=\textwidth]{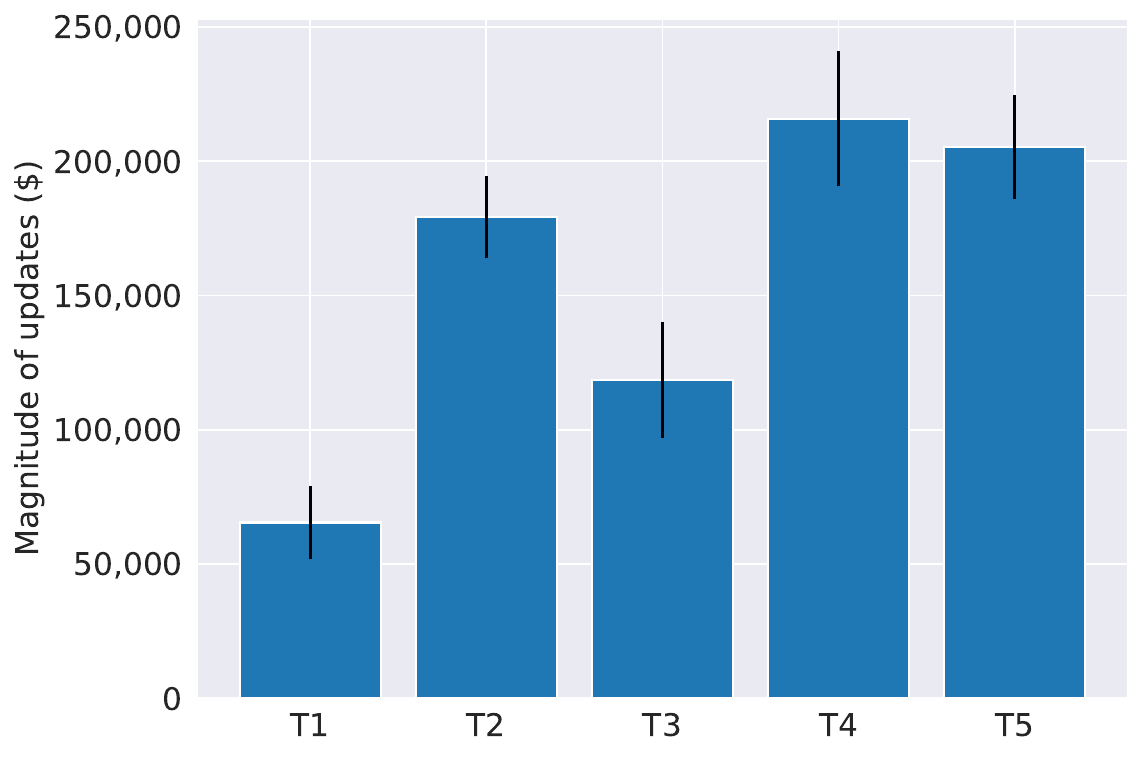}
         \caption{H1' magnitude: Magnitude of updates. The y-axis shows by how much the decisions shown in the review phase were updated.}
         \label{fig:H1_prime_magnitude}
    \end{subfigure}
    \hfill
    \caption{H1': Effect of the interventions on people's propensity to update decisions, across the subset of apartments that were shown in the review phase. The experimental conditions T1--T5 are shown on the x-axis. We report mean values calculated across respondents $\pm$ 1.96 standard errors of the mean (SEM).}
    \label{fig:H1_prime}
\end{figure}

\subsection{H1': Propensity to Change Particular Decisions} \label{subsec:resultsH1'}
H1 considers the overall effect of our interventions across all 30 apartments. However, in T3--T5 participants were able to update only a subset of their initial decisions. With H1' we account for this and focus on the effect of our interventions across the apartments shown in the review phase. 

In all five experimental conditions, respondents updated some of the decisions they were shown in the review phase. As in the analysis for H1, the number of decisions that were updated and the magnitude of the updates varied significantly between treatments. When compared to the baseline treatment T1, our interventions T2, T4 and T5 result in a higher propensity to update decisions for the particular apartments shown in the review phase. I.e., \textbf{our findings support H1'}. Again, this holds both for the number of decisions that were updated (H1' binary) and the magnitude of change (H1' magnitude). \looseness=-1

\xhdr{H1' binary: Number of Decisions Updated}
The second column of Table \ref{tab:regression} and Figure \ref{fig:H1_prime_binary} provide information about the fraction of apartments shown in the review phase that respondents updated. For treatments T1 and T2 the results are identical to those related to H1, since all 30 apartments were shown in the review phase. Namely, in T1 respondents updated $29\%$ of their decisions (8.7/30 apartments), while they updated $67\%$ of their decisions (20.2/30 apartments) in T2. For T3--T5, results change substantially once we account for the fact that respondents could not update all 30 apartments in the review phase. While T3 was not significantly different than T1 for H1, for H1' we identified a significant difference between these two treatments. Namely, respondents updated $44\%$ of the decisions they were shown in the review phase in T3. The effects of T4 and T5 were even stronger, with respondents updating $69\%$ and $74\%$ of decisions they had access to in the review phase. \looseness=-1

Our statistical analyses indicate that all treatments have significantly stronger effects than the baseline treatment T1 ($p<0.001$). Treatments T4 and T5 are not significantly different from each other ($p=0.1497$), but they both have stronger effects than T3 ($p<0.001$). Unlike for H1, T2 is not the treatment with the strongest effect. While it has a significantly stronger effect than T1 and T3 ($p<0.001$), it is not significantly different from T4, and it has a weaker effect than T5 ($p=0.0421$).

On a high level, we found that people are more likely to update their decisions when asked to review them as a series of pairwise comparisons and when they are provided with machine advice.

\xhdr{H1' magnitude: Magnitude of Updates}
The results of our analysis about the magnitude of changes are in line with our results about the amount of decisions that were updated (fourth column of Table \ref{tab:regression} and Figure \ref{fig:H1_prime_magnitude}). For T1 and T2, the results are the same as for H1: people update their decisions by approximately \$65,461 in T1 and by \$179,096 in T2. In T3--T5 the magnitude of the updates is significantly higher than we saw in the analysis of H1, with an average update close to \$117,719 in T3, \$206,157 in T4 and \$195,508 in T5.

T1 has a significantly weaker effect than the remaining four treatments ($p<0.001$), and T3 is in turn has a significantly weaker effect than the remaining three treatments ($p<0.001$), which are not significantly different between each other.

In other words, respondents updated their decisions by a larger amount when reviewing them as a series of pairwise comparisons and when they had access to machine advice.

\begin{figure}[t]
    \begin{subfigure}[t]{0.48\textwidth}
         \centering
         \includegraphics[width=\textwidth]{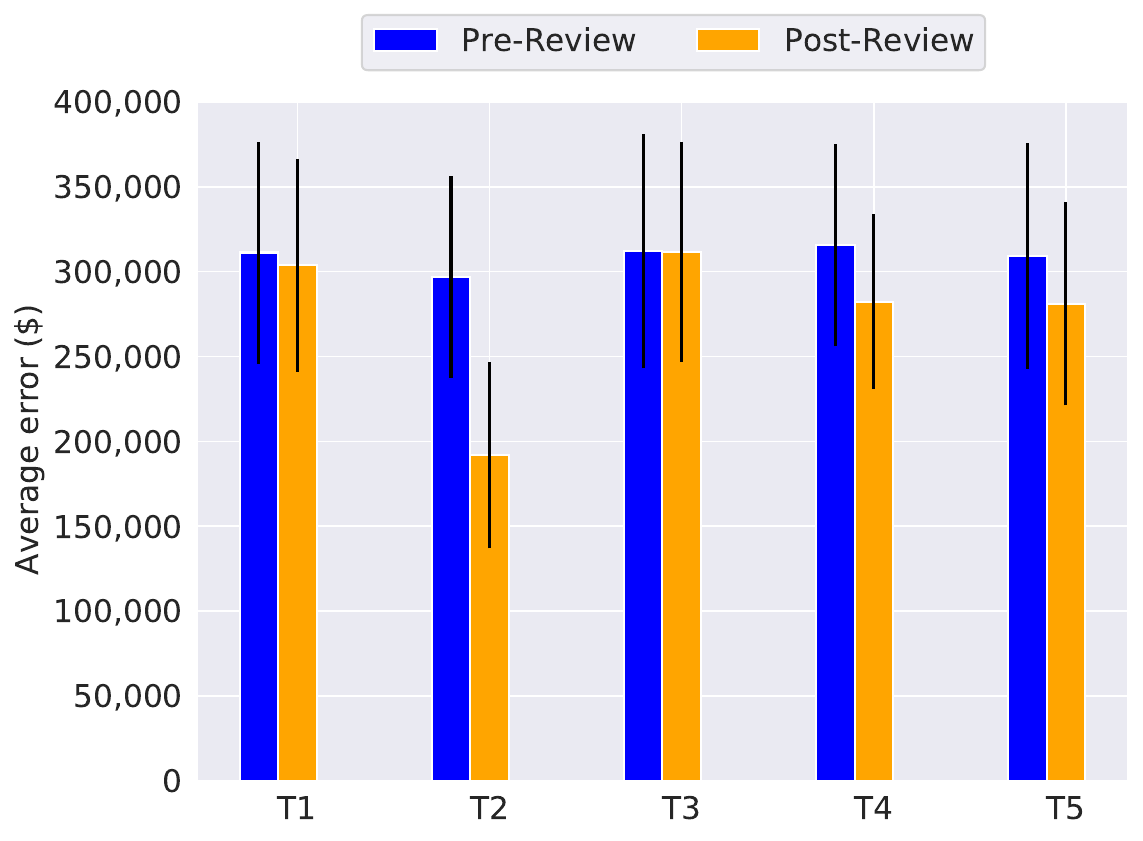}
         \caption{The y-axis shows the average error in people's decisions. The error in pre-review decisions is shown in blue, while the post-review error is shown in orange.}
         \label{fig:H2_raw}
    \end{subfigure}
    \hfill
    \begin{subfigure}[t]{0.48\textwidth}
         \centering
         \includegraphics[width=\textwidth]{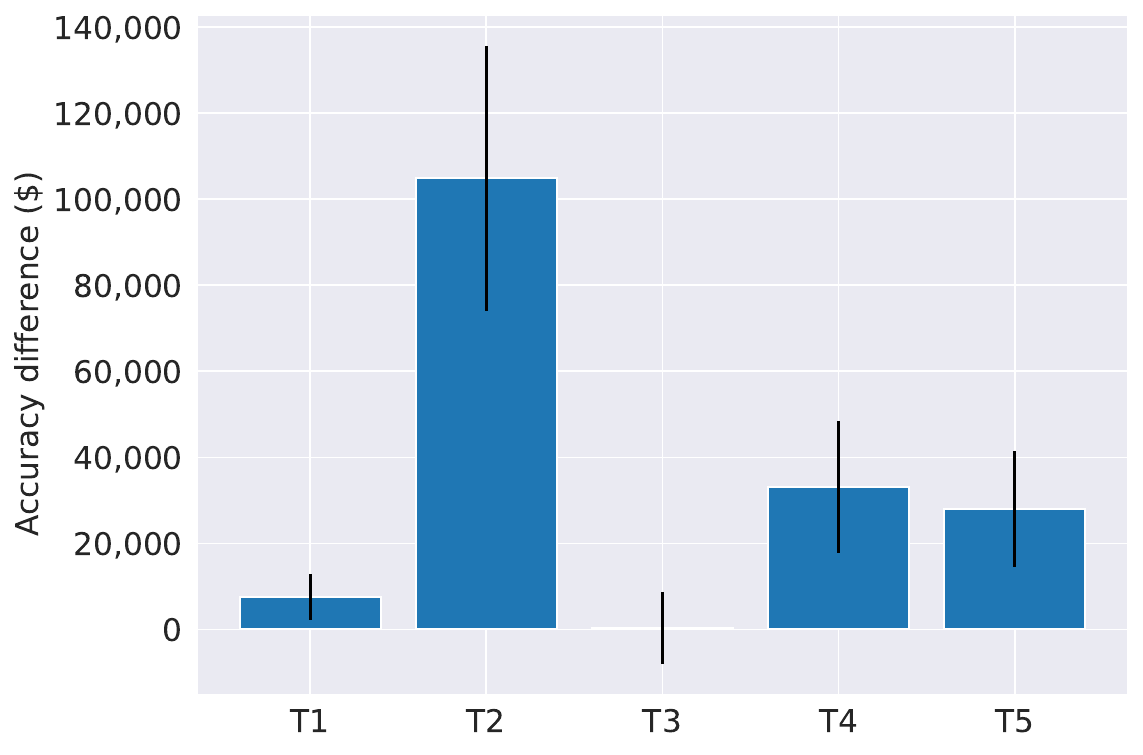}
         \caption{The y-axis shows the difference between the error in people's pre-review and post-review decisions.}
         \label{fig:H2}
    \end{subfigure}
    \hfill
    \caption{H2: Effect of the interventions on the accuracy of respondents' decisions. The experimental conditions T1--T5 are shown on the x-axis. We report mean values calculated across respondents $\pm$ 1.96 standard errors of the mean (SEM).}
    \label{fig:H2_full}
\end{figure}

\subsection{H2: Accuracy of Respondents' Decisions} \label{subsec:resultsH2}
Next, we study the impact of our interventions on the quality of the decisions---the accuracy of people's estimates. We find that interventions T2, T4 and T5 significantly improve the accuracy of people's post-review decisions, compared to the baseline treatment T1. That is, \textbf{our results are in line with H2}.

In H1 and H1' we found that all five of our experimental conditions influenced people's decisions. However, not all of the reviewing procedures led to an increase in the accuracy of people's decisions. As shown in the fifth column of Table \ref{tab:regression} and in line with Figure \ref{fig:H2_full}, the reviewing procedure utilized in T1 and T3 did not lead to a significant increase in the accuracy of people's post-review estimates, compared to their initial estimates. However, the remaining treatments did have a significant positive effect. Both T4 and T5 led to an increase in accuracy that is significantly higher than the one observed in T1 and T3 ($p<0.001$). In T4 people's estimates of apartment prices improved by an average of \$33,132, and in T5 by \$28,054. The difference between T4 and T5 was not significant ($p=1$). T2 led to a significantly higher improvement in accuracy ($p<0.001$) than the remaining treatments---people's post review estimates were closer to the ground truth by an average of \$104,863, compared to their initial estimates.

That is, while all treatments influenced people's decisions, not all of them led to an improvement in the accuracy of people's decisions. Only reviewing meaningfully selected pairs of apartments and having access to machine advice increased the accuracy of people's estimates.

In addition to this pre-registered confirmatory analysis, in Appendix \ref{subsec:appendix_accuracy} we present the results of additional exploratory analyses about the agreement between respondents' estimates and the ground truth. We find that the results presented in this subsection are robust with respect to an alternative measure of accuracy. We also offer descriptive insights about the directionality of people's errors. Namely, we find that our respondents systematically underestimate apartment prices in their initial estimates, and that the reviewing procedure reduces this bias in the directionality of respondents' errors.

\begin{figure}[t]
    \begin{subfigure}[t]{0.48\textwidth}
         \centering
         \includegraphics[width=\textwidth]{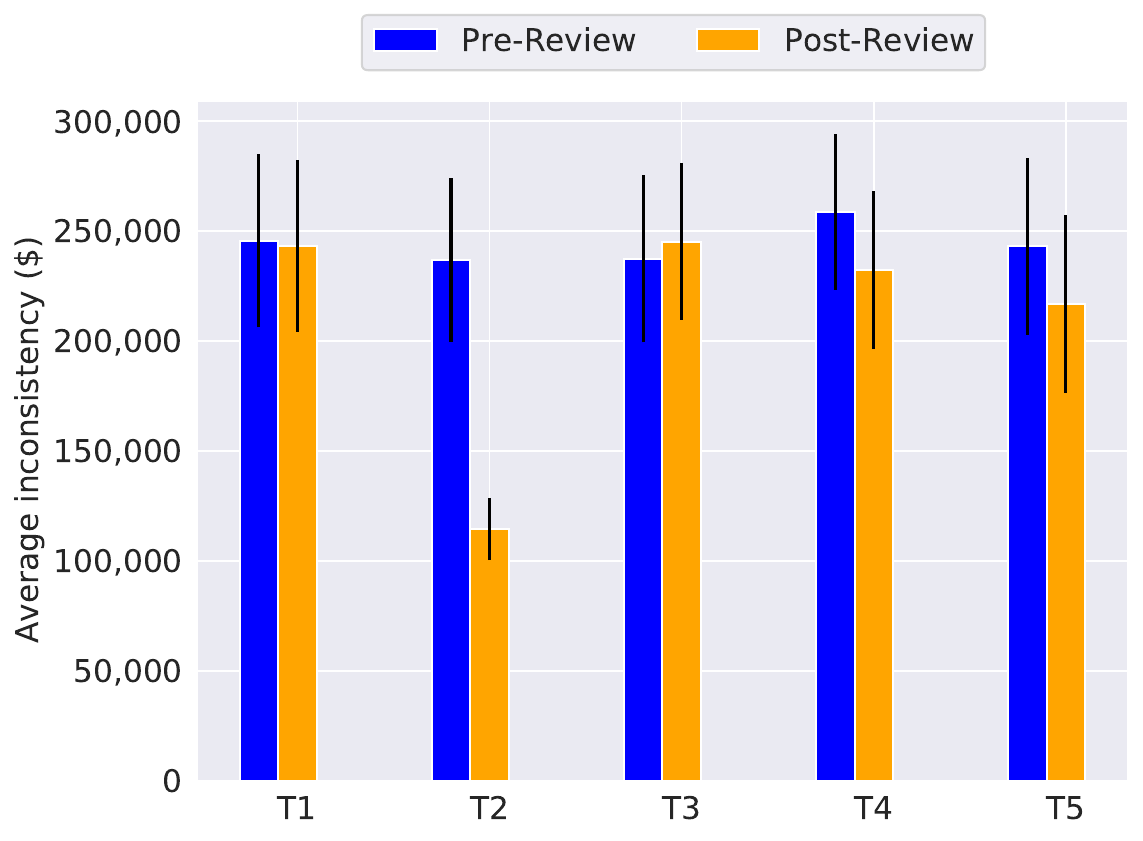}
         \caption{The y-axis shows the average inconsistency in people's decisions. The inconsistency in pre-review decisions is shown in blue, while the post-review inconsistency is shown in orange.}
         \label{fig:H3_raw}
    \end{subfigure}
    \hfill
    \begin{subfigure}[t]{0.48\textwidth}
         \centering
         \includegraphics[width=\textwidth]{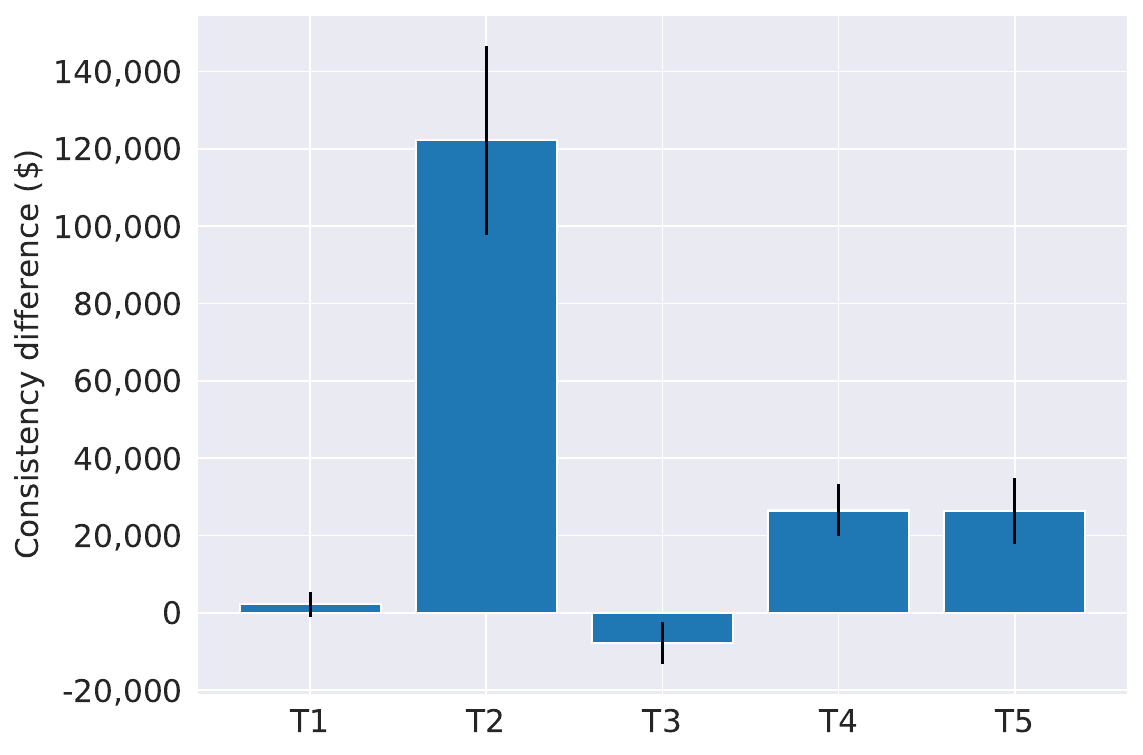}
         \caption{The y-axis shows the difference between the inconsistency in people's pre-review and post-review decisions.}
         \label{fig:H3}
    \end{subfigure}
    \hfill
    \caption{H3: Effect of the interventions on the consistency between respondents' decisions. The experimental conditions T1--T5 are shown on the x-axis. We report mean values calculated across respondents $\pm$ 1.96 standard errors of the mean (SEM).}
    \label{fig:H3_full}
\end{figure}

\subsection{H3: Consistency Between Respondents' Decisions} \label{subsec:resultsH3}
In this subsection, we investigate the effects of our interventions on the consistency between the decisions of different respondents. The patterns we identify are qualitatively similar to those related to the accuracy of people's decisions (H2). Namely, our interventions T2, T4 and T5 lead to a significantly higher increase in consistency between the post-review decisions of different respondents, compared to the control condition T1. That is, \textbf{the results support H3}. 

As shown in the sixth column of Table \ref{tab:regression} and in line with Figure \ref{fig:H3_full}, treatments T1 and T3 do not lead to an increase in people's consistency, while T2, T4 and T5 do. Compared to the consistency between respondents' pre-review estimates, T4 and T5 increase respondents' post-review consistency by \$26,486 and \$26,314 respectively. The difference between T4 and T5 is not significant ($p=1$), and the increase observed in both of these treatments is significantly higher than the effects observed T1 and T3 ($p<0.001$). Treatment T2 increases the degree of consistency in people's post-review decisions by an average of \$122,188, and this effect is significantly higher than the ones observed in any of the other treatments ($p<0.001$).

On a high-level, we found that while all treatments influenced the respondents' decisions, some of them did not have an impact on the degree of consistency between the estimates of different respondents. Comparisons of meaningfully selected pairs of apartments and access to machine advice have yet again proven to be effective strategies.

In addition to this pre-registered confirmatory analysis, we exploratively study a broader set of measures of consistency in Appendix \ref{subsec:appendix_consistency}. In short, we find that our results about the effects of our treatments on consistency are robust across a variety of measures of inter-annotator consistency.

\subsection{Exploratory Analysis: Agreement with Machine Advice} 
In Sections \ref{subsec:resultsH2} and \ref{subsec:resultsH3}, we studied the degree of agreement between our respondents' estimates and the \emph{ground truth} (Section \ref{subsec:resultsH2}), and the agreement between the estimates of \emph{different respondents} (Section \ref{subsec:resultsH3}). For treatments that utilize algorithmic assistance---that is, treatments T2, T4 and T5---we can compare our respondents' estimates to another variable of interest: \emph{machine advice}. In Appendix \ref{subsec:appendix_advice}, we report the details of an exploratory analysis of the agreement between people's estimates and the predictions made by the algorithmic decision aids they had access to. Here, we briefly summarize our findings.

At a high-level, we find that respondents take machine advice. When given the opportunity to review their decisions in T2, T4 and T5, respondents' update the majority of their estimates in a way that aligns with the decision aids' predictions---for example, increasing their estimate when the decision aid's prediction is higher in T2. Nevertheless, a nonnegligible fraction of the time, respondents choose not to update their estimates. Still, respondents rarely go against machine advice; few estimates are updated in the direction opposite of the decision aids' predictions.
\section{Discussion} \label{sec:discussion}

We conclude this work with a discussion about the implications (Subsection \ref{subsec:implications}) and limitations (Subsection \ref{subsec:limitations}) of our research, and a concluding summary of our work (Subsection \ref{subsec:conclusion}).

\subsection{Implications} \label{subsec:implications}

\xhdr{T2 - Traditional Machine Advice} Reviewing past decisions with access to machine advice has proven to be more effective than doing so without machine advice, not only in terms of people's propensity to update their decisions (H1), but also in terms of increasing the accuracy (H2) and consistency (H3) of their decisions. Our findings related to H1 and H2 are in line with recent research on machine-assisted decision-making and real world applications that have demonstrated that people are willing to take machine advice, particularly when the advice is highly accurate \cite{yin2019understanding} (as is the case in the real estate appraisal setting we consider). We contribute to this line of research by further demonstrating the effectiveness of traditional decision aids in increasing between-respondent consistency (H3).

Our results indicate that in settings where (i) one has access to ground truth data that enables the development of accurate decision aids, and (ii) it is deemed normatively desirable or acceptable to explicitly steer people towards making decisions in line with machine predictions, traditional algorithmic decision aids are an effective tool for doing so.

\xhdr{T3 - Randomly Selected Pairwise Comparisons} Past research in psychology \cite{stewart2005absolute, miller1956magical} and computer science \cite{narimanzadeh2023crowdsourcing} has found that people are better at making comparative judgments than absolute ones in certain contexts. Still, people's pairwise preferences are known to be inconsistent \cite{koczkodaj1993new, brunelli2015axiomatic, abel2018inconsistency}. Building upon both lines of research, we investigated if people's decisions may benefit from being reviewed in a series of randomly selected pairwise comparisons, instead of one-by-one. 

If this intervention had proven to be effective, it would have important design implications. This intervention would be suitable for low-resource environments, where it is difficult or impossible to develop machine decision aids, due to a lack of data for training them, the inherent difficulty of making accurate predictions in the decision-making task at hand, or a lack of well-established notions of objective ground truth. 

However, in our experiment, we did not find a significant difference between the accuracy and consistency of decisions that were reviewed one-by-one and those reviewed in randomly selected pairs. Hence, our results suggest that it might not be sufficient to switch from absolute to comparative decision-making when reviewing decisions, without carefully considering how one selects which pairs of decisions to review.

\xhdr{T4 - Consistency-Based Pairwise Comparisons} 
In T3, respondents are simply asked to review random pairs of decisions, but in T4 the reviewing procedure is guided by a machine decision aid. At first, this may not be evident---traditional decision aids (such as the one utilized in T2) usually attempt to predict correct decisions, and provide people with those predictions as advice. The decision aid used in T4 is quite different---it attempts to predict typical human decisions, and asks people to review their decisions when they do not match the predicted ones. That is, the algorithmic assistance in T4 helps people determine which pairs of decisions to review. This format of algorithmic assistance is---to the best of our knowledge---novel in the machine-assisted decision-making literature.

Prior research identified scenarios in which people's decision quality improves upon switching from absolute to comparative decision making \cite{stewart2005absolute, miller1956magical,narimanzadeh2023crowdsourcing}. We found that switching from an absolute to a comparative reviewing procedure is an effective strategy for increasing the accuracy and consistency of respondent's decisions \emph{only when} respondents compare \emph{meaningfully} selected pairs of inputs. The ineffectiveness of T3 demonstrates that a naive approach of randomly selecting pairs of decisions is not sufficient to reap the benefits of pairwise comparisons. In T4, our goal was to identify an approach for selecting pairs that would lead to an improvement in two metrics of interest---accuracy and between-respondent consistency. We have demonstrated that our approach succeeds in doing so.

While the decision aid utilized in T4 is less effective than the one from T2, it has two important advantages: (i) it does not require access to ground truth data, and (ii) it does not provide explicit advice on how to update decisions. The former makes this approach suitable for increasing accuracy and consistency even in environments where one does not have access to ground truth data, or for increasing consistency even in settings where the notions of ``ground truth'' and ``accuracy'' cannot be meaningfully defined. The latter makes it applicable even in settings where it is not normatively desirable to explicitly steer people towards specific decisions (e.g., due to concerns about silencing minority opinions), but to prompt people to review their own decisions and make them mutually consistent. That is, this intervention may be an excellent candidate for future research on \emph{intra}-annotator notions of consistency.

\xhdr{T5 - Consistency-Based Pairwise Comparisons with Advice} 
Most research on machine-assisted decision-making (e.g., \cite{grgic2019human, yin2019understanding, poursabzi2021manipulating}, reviewed in detail in Section \ref{sec:related_work}) focused on algorithmic decision aids that provide explicit advice to human decision makers. Given the plethora of evidence about the effectiveness of explicit advice, one may have expected T5 to be more effective than T4. However, perhaps surprisingly, T5 was not significantly more effective than T4 with respect to any of the dependent variables we studied. Reviewing past decisions as a series of meaningfully selected pairwise comparisons is equally effective with and without explicit machine advice. Hence, our results suggest that when it is deemed normatively undesirable to explicitly steer people towards specific response options, one can omit machine advice without impeding the effects of the reviewing procedure on the accuracy and consistency of human decisions.

This lack of a significant difference between T4 and T5 showcases the effectiveness of more subtle, implicit forms of algorithmic guidance. The algorithmic guidance in T4 is perhaps closer to a \emph{nudge} \cite{thaler2009nudge} than to decision aids in the judge-advisor system (JAS) paradigm \cite{bonaccio2006advice} typically studied in the machine-assisted decision-making literature. The effectiveness of nudging has been extensively studied in the social science literature \cite{thaler2009nudge}, but also in CS, particularly in HCI \cite{caraban201923} and research on recommender systems \cite{jesse2021digital}. Our results suggest that research on AI-assisted decision-making could also benefit from considering a broader set of algorithmic interventions, including implicit advice and subtle nudges.

In future research, it would be interesting to study why explicit machine advice does not have an effect in this setting. We hypothesize this might be caused by its redundancy: when comparing two apartments, respondents might be able to infer the majority's comparative valuation of these apartments. Research on incentive mechanisms that rely on people's ability to predict others' responses provides some backing to this hypothesis. Namely, peer prediction mechanisms \cite{miller2005eliciting}, in particular Bayesian Truth Serums and similar methods \cite{prelec2004bayesian, radanovic2013robust, radanovic2014incentives, krupka2013identifying}, ask respondents to predict what others will report in order to design proper incentives that incentivize truthful reporting. If people are able to accurately predict the majority's comparative valuations, explicit machine advice may not provide any additional information to respondents, and hence have no effect on their decisions. Future studies can test this hypothesis by evaluating people's ability to predict others' pairwise comparisons.

\subsection{Limitations and Future Work} \label{subsec:limitations}

\xhdr{Notions of Consistency}
In this paper, we explored the effects of our interventions on several measures of \emph{inter}-annotator consistency. In future work, it would be interesting to go beyond inter-annotator consistency, and consider notions of \emph{intra}-annotator consistency. 

The simplest extension would be the study of intra-annotator consistency \emph{across time}. That is, instead of measuring the degree of consistency between different respondents for the same input, one could measure the degree of consistency of the same annotator for the same input in different points in time. This extension requires minimal changes to our experimental design---namely, it requires conducting a longitudinal human-subject study. This line of work could provide important insights about moderating the effects of cognitive biases that lead to a person's inconsistency through time, such as dynamic inconsistency and hyperbolic discounting \cite{loewenstein1992anomalies, thaler1981some}, or the (contested) ``hungry judge'' effect \cite{danziger2011extraneous}.\footnote{While the ``hungry judge'' effect \cite{danziger2011extraneous} is often referenced as an argument in favor of introducing algorithmic assistance in legal decision making, the validity of the study's findings has been much debated in recent literature \mbox{\cite{glockner2016irrational, chatziathanasiou2022beware}}.}

Intra-annotator consistency across time is closely related to counterfactual questions such as ``Would the decision-maker have made the same decision in a different point in time?'' A different notion of consistency---intra-annotator consistency \emph{across inputs}---addresses the question ``Does the decision-maker make similar decisions for similar inputs?'' This line of research is closely related to research on individual fairness. 

A central problem in studying both intra-annotator consistency across inputs and individual fairness lies in defining the similarity metric which determines which inputs should be treated as similar. Prior work on individual fairness has assumed such similarity metrics to be given \cite{dwork2012fairness}, or defined them based on the inputs' ground truth labels \cite{joseph2016fairness, liu2017calibrated}, distance in transformed feature spaces that align with certain distributive fairness criteria \cite{zemel2013learning, lahoti2019ifair}, or---as we implicitly did in T4 and T5---based on human judgments about input similarity \cite{ilvento2019metric, wang2019empirical, lahoti2019operationalizing, jung2019algorithmic}. As a promising direction for future work, we highlight the study of intra-annotator consistency across inputs with personalized similarity metrics, i.e., the development of methods for identifying decisions that are outliers, inconsistent with the other decisions made by the same respondent.

In this work we study methods for alleviating inter-annotator inconsistency. Deciding which notion of inconsistency is appropriate to apply in a given setting is inherently a normative question. Hence, we invite future work not only on formalizing and operationalizing different notions of inconsistency, but also philosophical and policy discussions on the desirability of different---and as discussed below, any---notions of consistency in specific settings.

\xhdr{Benefits of Human Inconsistency}
This paper focused on settings where inconsistency between multiple decision-makers may be deemed undesirable. As such, the proposed methods are not applicable and should not be applied in settings where diversity in people's beliefs, perceptions, and behavior may be beneficial, or considered normatively desirable.

Diversity in people's decisions may reflect the differences in their skill set and background knowledge, and these differences can be exploited to improve decision-making quality \cite{welinder2010multidimensional}. Diversity in the composition of groups increases the diversity in the problem solutions that team members propose, which in turn increases the quality of group decisions \cite{wanous1986solution}. Heterogeneity in teams can benefit group performance, since the diversity in the perspectives of different team members can foster creativity and innovation \cite{roberge2010recognizing}. 

Furthermore, people's beliefs, perceptions and behavior are known to correlate with their socio-demographics and life experiences. For instance, the rich literature on Moral Foundations Theory found that sociodemographic characteristics such as political views \cite{graham2013moral,graham2011mapping}, gender \cite{graham2011mapping}, and educational attainment \cite{van2014moral} correlate with people's moral views. Lived experiences, such as growing up during an economic recession \cite{giuliano2014growing} or experiencing economic shocks \cite{alesina2011preferences, margalit2019political}, correlate with people's preferences regarding social policies. Socio-demographic factors \cite{grgic2022dimensions, albach2021role, pierson2017demographics} and people's life experiences \cite{grgic2022dimensions} were also found to correlate with people's moral judgments about algorithmic fairness. That is, the decisions made by members of minority groups may systematically differ from those made by members of the majority. Therefore, methods for reducing inconsistency between people may---inadvertently or on purpose---explicitly steer people's decisions toward the majority's view, thereby silencing the minorities' views. \looseness=-1

Hence, prior to applying any methods for reducing inconsistency between decision-makers, it is crucial to understand the sources of this variation, and to evaluate whether reducing it would be appropriate and normatively desirable in the decision-making task at hand.

\xhdr{Generalizability to Other Domains}
In this work, we focus on a real estate appraisal scenario. While we find large and statistically significant effects of our interventions for the task at hand, we invite future work that will systematically explore which types of scenarios our findings generalize to. We opted for this scenario because many laypeople have prior experience with property valuation (e.g., searching for, purchasing or selling real estate), but most laypeople do not make highly accurate estimates of real estate prices. The task we considered may be in the sweet spot between too difficult and too easy for our respondent sample. We hypothesize that our findings may not generalize to tasks on either of the extremes.

For tasks that people find easy, such as visual recognition tasks, interventions may not have an effect if people already exhibit high degrees of accuracy and consistency, hence not allowing room for significant improvement along either dimension. For tasks that people find difficult, such as criminal risk prediction, both people and algorithms may exhibit low levels of accuracy. For instance, in a pilot study we conducted using the ProPublica COMPAS dataset \cite{propublica_story}, algorithmic advice (T2, with an accuracy of 58\%) did not have a significant impact on consistency since it increased consistency for some cases, while decreasing it for others. The latter typically occurred for the non-negligible number of cases where respondents initially made correct predictions, but incorrect machine advice steered them away from their initial responses, decreasing their accuracy and consistency levels.

We further note that we studied the effects of algorithmic assistance on respondents' accuracy and consistency in a task where (i) the notion of ground truth is well-defined, and (ii) one could deem consistency between professionals to be normatively desirable. However, the notions of accuracy and consistency we studied may not be suitable for every decision-making task. For tasks where there is no well-defined notion of ground truth, such as subjective tasks, the notion of accuracy cannot be well-defined either. For subjective tasks, it may also be deemed normatively undesirable to promote inter-annotator consistency, and one may be in favor of intra-annotator consistency, or a different metric instead. In short, the problem of choosing an appropriate evaluation metric is a policy question that requires an understanding of the underlying normative goals of utilizing algorithmic assistance in the decision-making task at hand.

\xhdr{Interventions}
In this paper, we report the effects of five different interventions. In pilot studies we considered one additional intervention, where we asked respondents to review all of their initial estimates on the same page, sorted by the apartment prices they estimated. We initially conjectured that this may allow respondents to conduct comparisons of apartments that they deemed to have similar prices. However, since (i) this approach was not scalable to a large number of decisions, and (ii) the effects of this treatment showed no statistically significant difference from T1 and T3 in our pilots, we omitted it from our main study for brevity. 

We invite future work that would explore an even broader set of interventions. As reviewed in Section \ref{sec:related_work}, prior work on human advice taking behavior \cite{bonaccio2006advice} and on machine-assisted decision-making has identified numerous factors that influence how people take advice, including the decision aid's accuracy \cite{yin2019understanding}, explainability \cite{poursabzi2021manipulating}, and the stakes associated with the decision-making task \cite{grgic2019human}, and future work could incorporate some of these factors in their interventions. Future work could also build upon T4 and T5 by developing decision aids that not only predict which of two apartments is perceived as more expensive, but also identify apartments that are perceived to be equally expensive. Identifying data points that are perceived as deserving of similar outputs may be interesting not only for the study of noise in human decisions, but also for research on individual fairness.

\xhdr{Respondent Samples}
In our experiment we recruited a large and demographically diverse set of laypeople from the US. Future work could explore if our findings replicate in other cultures beyond the US. Additionally, it is worth noting that our sample consisted of laypeople, and it is possible that expert judgments of professionals such as real estate agents systematically differ from the perceptions of our lay sample. For instance, professionals may be substantially more accurate in their predictions, thereby having fewer opportunities to benefit from algorithmic advice. Hence, it may be interesting to replicate our experiment with industry professionals.

\subsection{Conclusion} \label{subsec:conclusion}
In this work, we studied methods for alleviating inconsistency in human decision-making. We identified several approaches that effectively influence human decisions, improving their accuracy and consistency with other respondents. We identified methods that are applicable to a wide variety of scenarios, including for settings where one has access to ground truth data for training decision aids (T2), as well as for settings where one only has access to human annotations (T4 and T5), but none for settings where no data is available (T3). All of the treatments that significantly improved decision accuracy and consistency relied on algorithmic assistance, be it explicit (T2 and T5) or implicit (T4). As a promising avenue for future work, we see the study of a broader set of notions of inconsistency, including intra-annotator consistency.

\bibliographystyle{ACM-Reference-Format}
\bibliography{denoise}

\pagebreak
\appendix
\section{Appendix} \label{sec:appendix}

In the main paper, we presented the results of our pre-registered confirmatory statistical analyses. In this appendix, we provide more details about our data and present the results of additional exploratory analyses. 

In Appendix \ref{subsec:appendix_details}, we provide more details about the effect sizes associated with the pre-registered hypotheses H1--H3. In Appendix \ref{subsec:appendix_accuracy} and \ref{subsec:appendix_consistency}, we present supplementary results related to H2 and H3 respectively. Namely, in Appendix \ref{subsec:appendix_accuracy} we explore the agreement between people's responses and \emph{ground truth} labels, while in Appendix \ref{subsec:appendix_consistency} we explore the agreement between \emph{different people's responses}. Finally, in \ref{subsec:appendix_advice} we explore the agreement between people's responses and \emph{machine advice}.

\subsection{Effect Sizes} \label{subsec:appendix_details}

When determining our study sample size, we aimed to recruit sufficiently many participants to detect at least medium-sized effects (Cohen's d = 0.5) at the significance level of 0.05 with 0.95 power, as detailed in our pre-registration, which can be found on the following url: \url{https://aspredicted.org/D7X_NKL}. 

In Table \ref{tab:cohens_d}, we report the values of Cohen's d calculated on our dataset. We find that treatment T3 is associated with the smallest effect sizes, followed by treatments T4 and T5, and finally T2---which has the largest effect sizes. When we compare the Cohen's d values in Table \ref{tab:cohens_d} with the regression coefficients reported in Table \ref{tab:regression}, we observe that our regression identified even small effects. Namely, for values of Cohen's d > 0.1, the corresponding regression coefficients are significantly different from 0 in the regression analysis.

\begin{table}[h]
\centering
\def\sym#1{\ifmmode^\textbf{#1}\else\(^\textbf{#1}\)\fi}
\begin{tabular}{l*{6}{c}}
    \toprule
        & \textbf{H1}: change, &  \textbf{H1'}: change, &  \textbf{H1}: change, &  \textbf{H1'}: change, &  \textbf{H2}: &  \textbf{H3}:\\
    &  bin., overall &  bin., specific &  mag., overall &  mag., specific &  accuracy &  consistency \\
    \hline \\
    [-0.75em]
     \textbf{T2} & 0.83 & 0.83 & 0.58 & 0.58 & 0.53 & 0.7\\
     \textbf{T3} & 0.004 & 0.33 & 0.07 & 0.28 & 0.05 & 0.07\\
     \textbf{T4} & 0.22 & 0.88 & 0.28 & 0.73 & 0.15 & 0.15\\
     \textbf{T5} & 0.29 & 1.02 & 0.27 & 0.72 & 0.13 & 0.16\\
    \bottomrule
\end{tabular}
\caption{Cohen's d. A value in row \emph{i} and column \emph{j} corresponds to the effect size of treatment \emph{i} on the variable \emph{j}. On a high-level, Cohen's d quantifies the difference between the means of variable \emph{j} for two groups of respondents: those assigned to treatment T1 and those assigned to treatment \emph{j}. More precisely, we report the values of Cohen's d as defined in \citet{cohen1988statistical}, calculated using the \texttt{esize} command in Stata, specifying the option \texttt{unequal}, to specify that the two groups should not be assumed to have equal variances.}
\label{tab:cohens_d}
\end{table}

\subsection{Agreement with Ground Truth} \label{subsec:appendix_accuracy}
In Section~\ref{sec:results}, we quantified accuracy as the degree of agreement between respondents' estimates and the ground truth. In this section, we explore if our results hold for a broader set of measures of accuracy. We further explore the directionality of the disagreement between people's estimates and the ground truth. That is, we investigate whether people tend to overestimate or underestimate apartment prices.

\begin{figure}[t]
    \centering
    
    \includegraphics[width=0.48\textwidth]
    {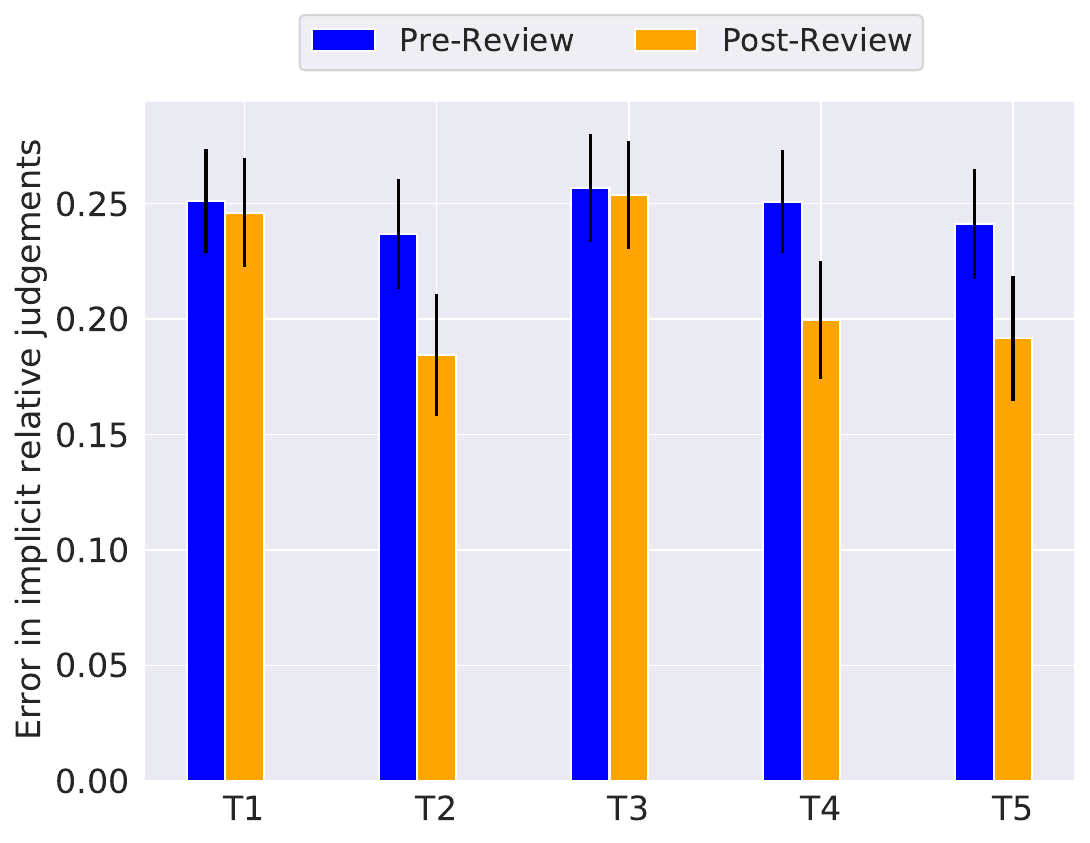}
         \caption{Error in people's implicit relative judgments. The y-axis shows the fraction of instances where people's implicit relative ordering of apartments (>,< or =) did not match the ground truth ordering based on the listing price. We report mean values calculated across all respondents and pairs of apartments $\pm$ 1.96 standard errors of the mean (SEM).}
         \label{fig:relative_accuracy}
    
\end{figure}

\subsubsection{Accuracy of People's Implicit Relative Judgments}\hfill\\
In this section, we go beyond the accuracy of people's absolute judgments about apartment prices, and consider the accuracy of their implicit relative judgments. We start by deriving people's implicit relative judgments from their absolute estimates. For each pair of apartments (A,B), we check if a respondent estimated Apartment A to be more expensive (>), less expensive (<) or equally as expensive (=) as Apartment B. Then we compare these implicit relative judgments with the ground truth (i.e., with the relative ordering of apartments based on their listing price). 

In Figure~\ref{fig:relative_accuracy}, we report the fraction of instances where people's implicit relative ordering differed from the apartments' true ordering. Descriptively, we observe that the error in people's pre-review relative estimates is similar across all experimental conditions. In experimental conditions T1 and T3, the error in people's pre- and post-review estimates remained similar. However, in T2, T4 and T5, the error decreased by 5.2, 5.1 and 5.0 percentage points respectively. That is, treatments T2, T4 and T5 reduced the error in people's implicit relative judgments by 22.1\%, 20.4\% and 20.5\% respectively. These exploratory findings are in line with our findings related to people's absolute judgments. \looseness=-1

\subsubsection{Directionality of Errors}\label{subsubsec:error_direction}\hfill\\
In this section, we take a closer look at the directionality of people's errors. Namely, we explore whether respondents tend to systematically overestimate or underestimate apartment prices.

\begin{figure}[t]
    \begin{subfigure}[t]{0.48\textwidth}
         \centering
         \includegraphics[width=\textwidth]{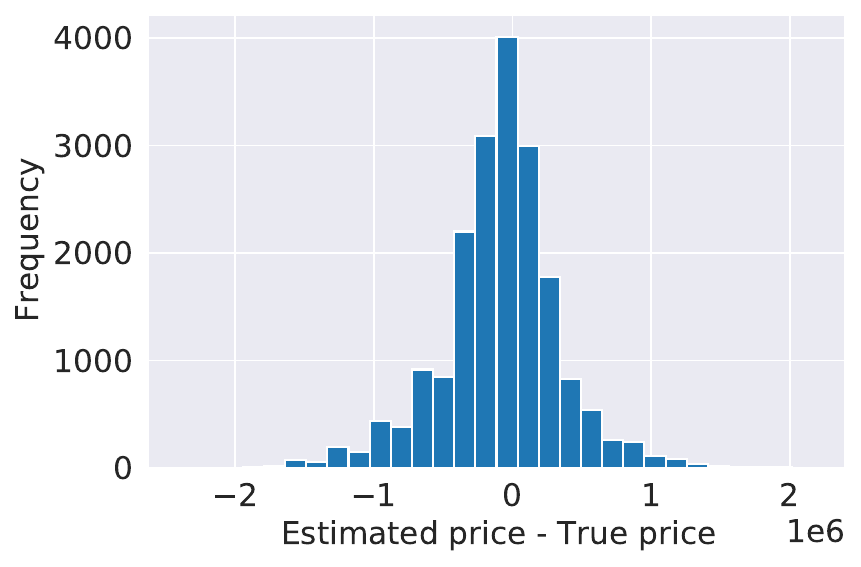}
         \caption{Distribution of errors in respondents' pre-review estimates.}
         \label{fig:pre_advice_distribution}
    \end{subfigure}
    \hfill
    \begin{subfigure}[t]{0.48\textwidth}
         \centering
         \includegraphics[width=\textwidth]{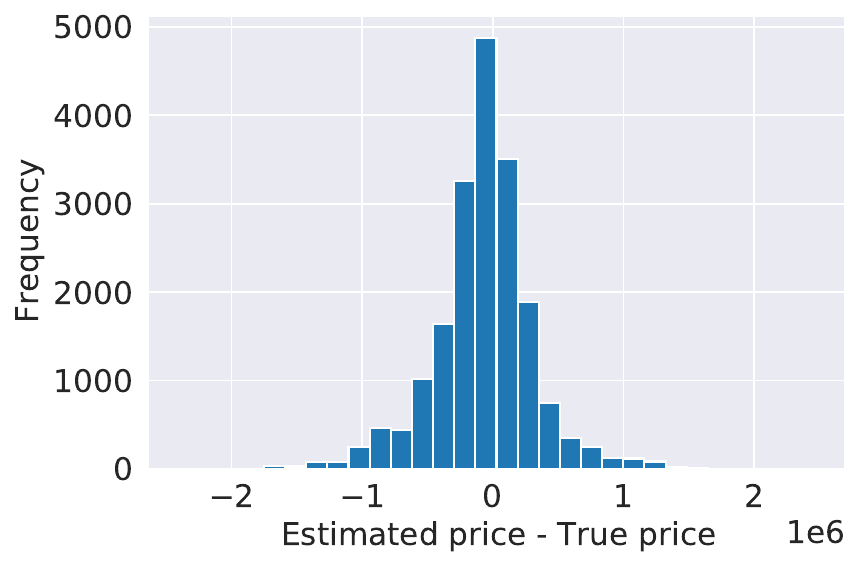}
         \caption{Distribution of errors in respondents' post-review estimates.}
         \label{fig:post_advice_distribution}
    \end{subfigure}
    \hfill
    \caption{Distribution of errors in respondents' estimates, across all treatments. The x-axis shows the magnitude of errors, i.e., the difference between the respondents' estimates and the apartments' true prices. The y-axis shows the number of responses in our dataset that exhibited a certain magnitude of error.}
    \label{fig:distribution_of_errors}
\end{figure}

\begin{figure}[t]
    \centering
    \includegraphics[width=0.48\textwidth]{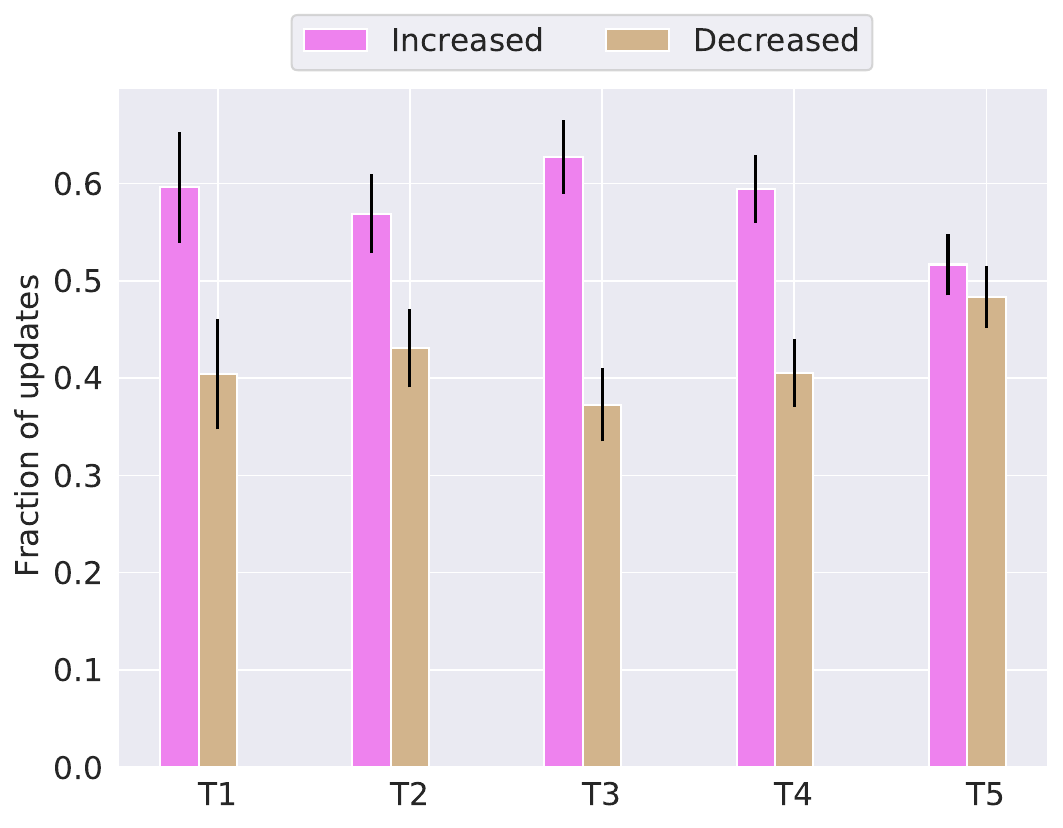}
    \caption{Directionality of response updates. The y-axis shows the fraction of revised responses that were updated to increase (blue) or to decrease (orange) the initial price estimates, for each of the experimental conditions T1-T5, shown on the x-axis.}
    \label{fig:post_advice_direction}
\end{figure}

In our experiment, we utilized a dataset of apartments introduced by \citet{poursabzi2021manipulating}. The dataset contains information about apartments located in New York City, that were listed for sale between 2013 and 2015. Since the prices of real-estate have increased between the mid-2010s and today, it is possible that people systematically overestimate the prices of these apartments. On the other hand, since we utilize a dataset of apartments located in New York City, which is significantly more expensive than other US cities, it is possible that people systematically underestimate the prices of these apartments. In this section, we investigate if people's errors exhibit either of these patterns. We note that even if people exhibited such a systematic bias in their errors, this would not affect the validity of our results. Any systematic overestimation or underestimation related to the apartments used as stimulus material would be the same across treatments, since they utilize the same 30 apartments. Since our hypotheses concern the differences between experimental conditions, this would not impact our results.

We commence by exploring the directionality of errors in people's initial estimates of apartment prices. In Figure \ref{fig:pre_advice_distribution} we show the distribution of errors in people's pre-review estimates. Values on the left side of the x-axis correspond to instances where respondents underestimated apartment prices, while values on the right correspond to instances of overestimating apartment prices. Descriptively, the distribution is fairly close to normal, but it exhibits a left skewness. That is, our respondents both underestimated and overestimated apartment prices in their initial estimates, but they were more likely to underestimate them. Next we investigate how participants updated their initial responses in our five experimental conditions. In Figure \ref{fig:post_advice_direction} we report the direction in which respondents updated their estimates in the review phase. When reviewing their responses, respondents were found to both increase and decrease their estimates of apartment prices, but they were were more likely to increase them. That is, respondents were more likely to initially underestimate apartment prices, but they were more likely to increase their initial estimates while reviewing them. This leads us to the distribution of errors in people's post-review estimates shown in Figure \ref{fig:post_advice_distribution}. Descriptively, the distribution of post-review errors is more narrow than the distribution of pre-review errors, in line with the increase in the accuracy of people's decisions. Additionally, the distribution is less skewed. That is, the reviewing procedure helped people reduce the systematic bias in the directionality of their errors.

\begin{figure}[t]
    \begin{subfigure}[t]{0.48\textwidth}
         \centering
         \includegraphics[width=\textwidth]{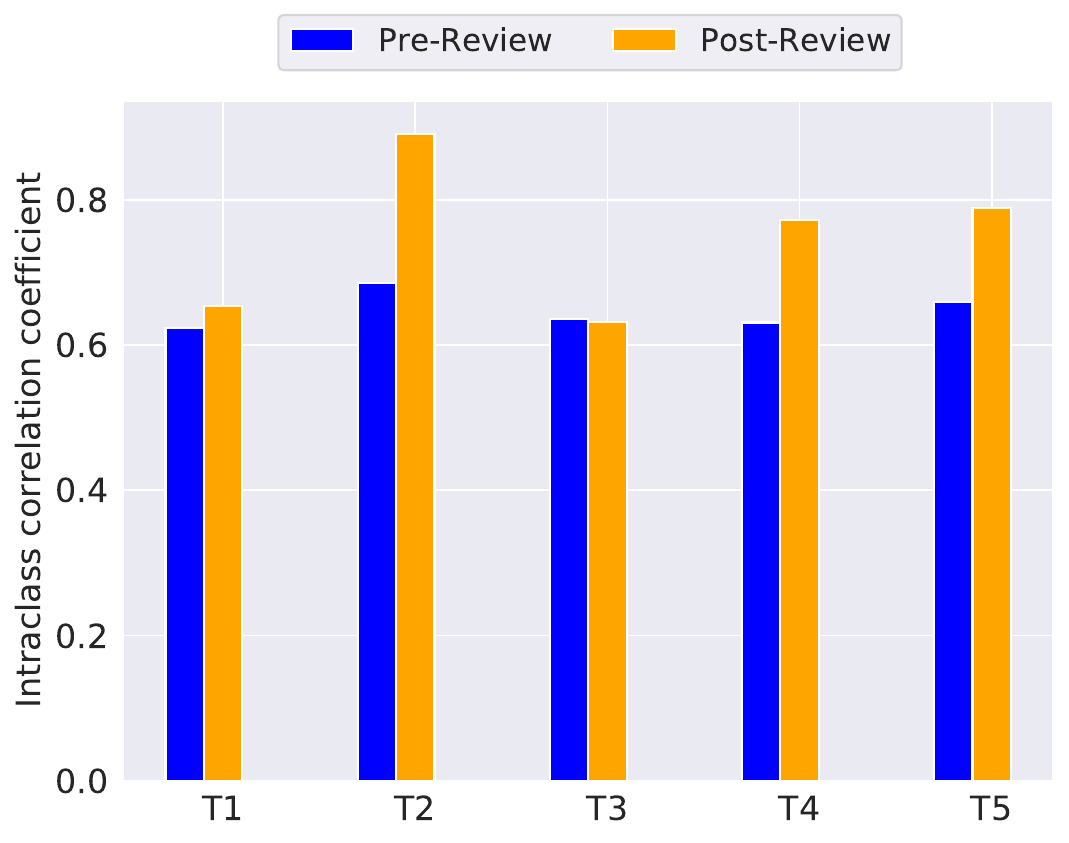}
         \caption{Intraclass correlation. The y-axis shows the Intraclass correlation among respondents over all the apartments. A higher value indicates a higher degree of consistency.}
         \label{fig:ICC}
    \end{subfigure}
    \hfill
    \begin{subfigure}[t]{0.49\textwidth}
         \centering
         \includegraphics[width=\textwidth]{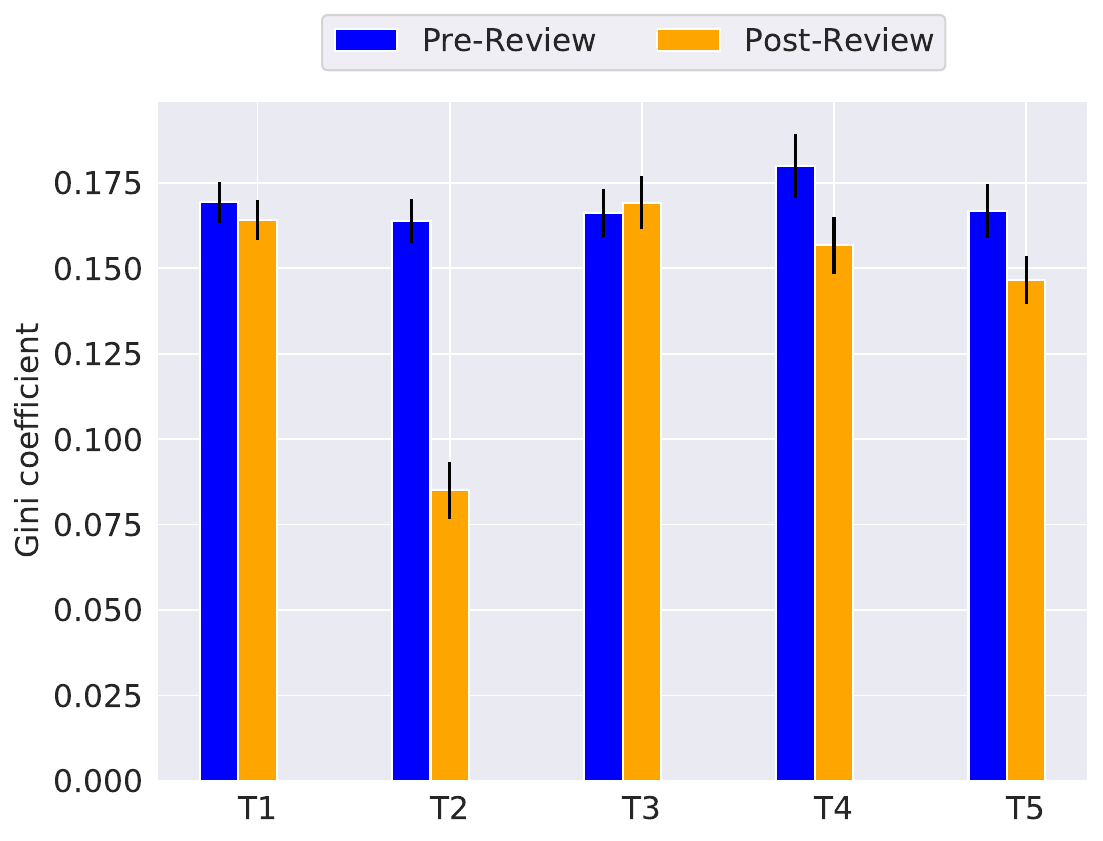}
         \caption{Gini coefficient. The y-axis shows the Gini coefficient of people's responses averaged across apartments. A lower value indicates a higher degree of consistency. We report mean values $\pm$ 1.96 standard errors of the mean (SEM).}
         \label{fig:gini}
    \end{subfigure}
    \hfill
    \caption{Effect of the interventions on the consistency of people's absolute judgments. The experimental conditions T1--T5 are shown on the x-axis.}
    \label{fig:conistency_metrics}
\end{figure}

\subsection{Agreement with Other Respondents} \label{subsec:appendix_consistency}
In Section~\ref{sec:results}, we quantified consistency as the degree of agreement between respondents’ estimates. In this section, we investigate whether our findings hold for a broader set of measures of inter-annotator consistency.

We explore the consistency of two types of dependent variables: the respondents' \emph{absolute} judgments, and their implicit \emph{relative} judgments. In the former, we quantify the consistency between respondents' estimates of apartment prices. In the latter, we focus on the consistency of respondents' judgements regarding the relative ranking or ordering of the apartments. That is, we evaluate whether different respondents assign similar relative positions to the apartments. 

We find that our results about the effects of the studied interventions on inter-annotator consistency are robust across a variety of measures. Descriptively, we observe that treatments T1 and T3 do not impact the degree of consistency between respondents, neither in terms of their absolute judgments nor in terms of their implicit relative judgments. On the other hand, treatments T2, T4 and T5 are found to improve both types of consistency notions. For people's absolute judgments, T2 leads to the greatest increase in consistency. For implicit relative judgments, T4 and T5 increase the overall ranking consistency the most, while all three treatments lead to a similarly large increase in pairwise ranking consistency.

\subsubsection{Consistency of People's Absolute Judgments}\hfill\\
\xhdr{Intraclass Correlation Coefficient (ICC)} The ICC is typically used as a metric to assess annotators' reliability. An estimate by annotator $i$ for apartment $a$ is modelled as $x_{i,a} = \mu + \alpha_i + \beta_a + \epsilon_{i,a}$, where $\mu$ is the unobserved overall mean, $\alpha_i$ models the random effect specific to annotator $i$, $\beta_{a}$ represents the random effect due to the features of apartment $a$, while $\epsilon_{i,a}$ represents the noise. With this model of annotator estimates, ICC is defined as follows:
\begin{equation}\label{eq:icc}
ICC = \frac{\sigma_{\beta}}{\sigma_{\alpha} + \sigma_{\beta} + \sigma_{\epsilon}}. 
\end{equation}
Here, $\sigma_{\beta}$ represents the variability in the estimates due to differences in the features of the apartments such as their size or number of rooms. $\sigma_{\alpha}$ captures the variability resulting from the differences in the scales used by different respondents, e.g., some respondents may consistently provide higher estimates than others. $\sigma_{\epsilon}$ accounts for the variability arising due to noise in respondents' evaluations. We calculated the ICC using the \texttt{Pingouin} library \cite{Vallat2018}. We report the ICC3 values, which---in line with our setting---assume a fixed set of $k$ respondents for each instance.

If the variability in the estimates is predominantly due to apartments' features the ICC value would be high. Conversely, if there is a high variance in the magnitude of the estimates due to differences in scales ($\sigma_\alpha$) or noise ($\sigma_\epsilon$) ICC would be lower. Essentially, the ICC captures how responses cluster for each apartment. A value of zero implies that there are no clusters, and each response is likely to be independent. A value of one implies that all the responses are the same.

Descriptively, we observe that the ICC of people's pre-review estimates is similar across all five treatments, as shown in Figure~\ref{fig:ICC}. In treatments T1 and T3 the ICC of people's post-review estimates remained similar to the ICC of their pre-review estimates. However, in treatments T2, T4 and T5 people's post-review estimates exhibited a higher ICC than their pre-review estimates. Namely, we observe an increase of 0.21, 0.14 and 0.13 in T2, T4 and T5 respectively. 

It is important to note that although ICC is a consistent statistic, it has a positive bias, i.e., it overestimates the true value. Additionally, it relies on several assumption such as $\alpha$, $\beta$ and $\epsilon$ having an expected value of zero and $\beta$ being uncorrelated with $\alpha$ and $\epsilon$. Below, we consider a metric that does not rely on such modeling assumptions: the Gini coefficient.

\xhdr{Gini Coefficient} The Gini coefficient is a measure of dispersion commonly used to quantify inequality within groups, such as wealth inequality within a nation. Unlike the ICC, the Gini coefficient directly focuses on the differences in respondents' estimates for each apartment, without any modeling assumptions. It is defined as follows:

\begin{equation}\label{eq:gini}
    G =  \frac{\sum_i^k \sum_j^k | x_i - x_j|}{2 \cdot k\cdot \sum_j x_j},
\end{equation}
where $x_*$ denotes the estimate by respondent $*$ and $k$ is the number of respondents. This value captures the dispersion of people's responses for a given apartment. A value of zero indicates that the responses are closely clustered together, while a value of one means the estimates are completely dispersed. To quantify the dispersion across all apartments, we calculate the average Gini coefficient across all 30 apartments.

We report our findings in Figure~\ref{fig:gini}. Descriptively, we found a similar trend as for the ICC. For treatments T1 and T3, the Gini coefficients of people's pre- and post-review estimates remained similar. On the other hand, in treatments T2, T4 and T5, people's post-review estimates exhibit a lower Gini coefficient, with a decrease of 0.08 in T2, and a decrease of 0.02 in T4 and T5.

\begin{figure}[t]
    \begin{subfigure}[t]{0.49\textwidth}
         \centering
         \includegraphics[width=\textwidth]{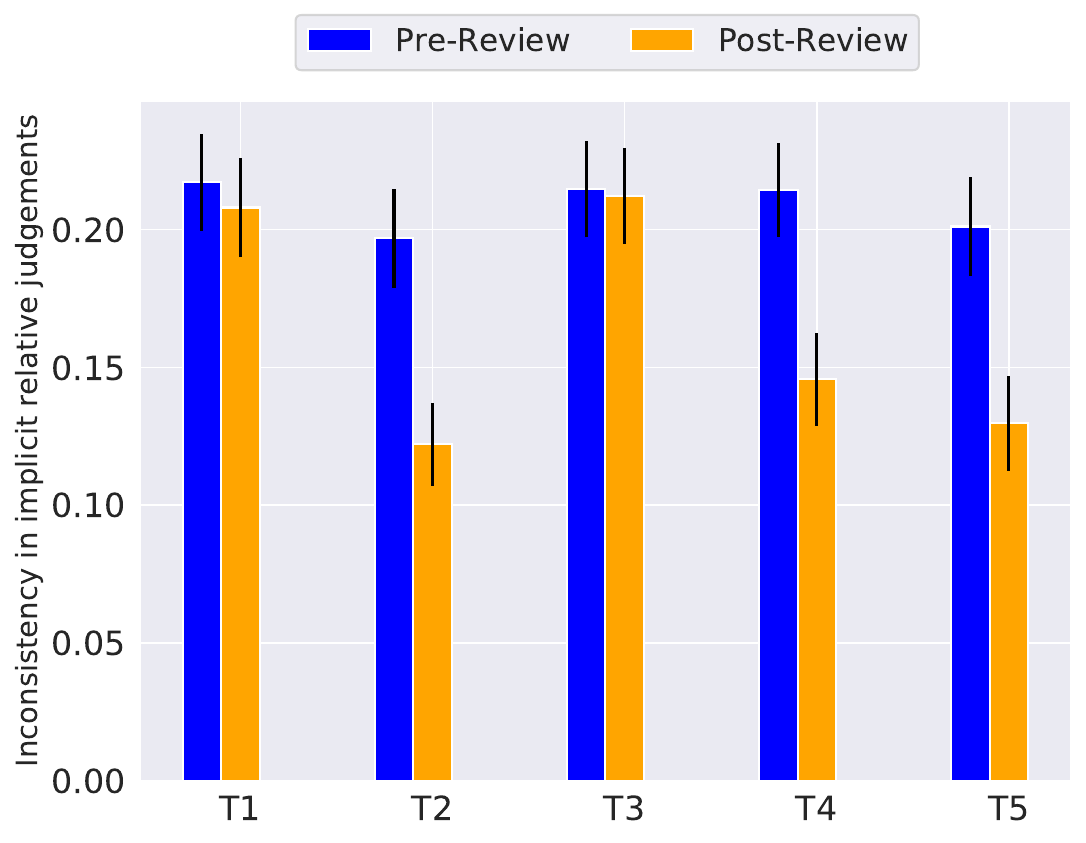}
         \caption{Inconsistency in people's implicit relative judgments. The y-axis shows the fraction of instances where people's implicit relative ordering of apartments (>,< or =) did not match the relative ordering of the majority of respondents. A lower value indicates a higher degree of consistency. We report mean values $\pm$ 1.96 standard errors of the mean (SEM).}
         \label{fig:majority_disagreement}
    \end{subfigure}
    \hfill
    \begin{subfigure}[t]{0.49\textwidth}
         \centering
         \includegraphics[width=\textwidth]{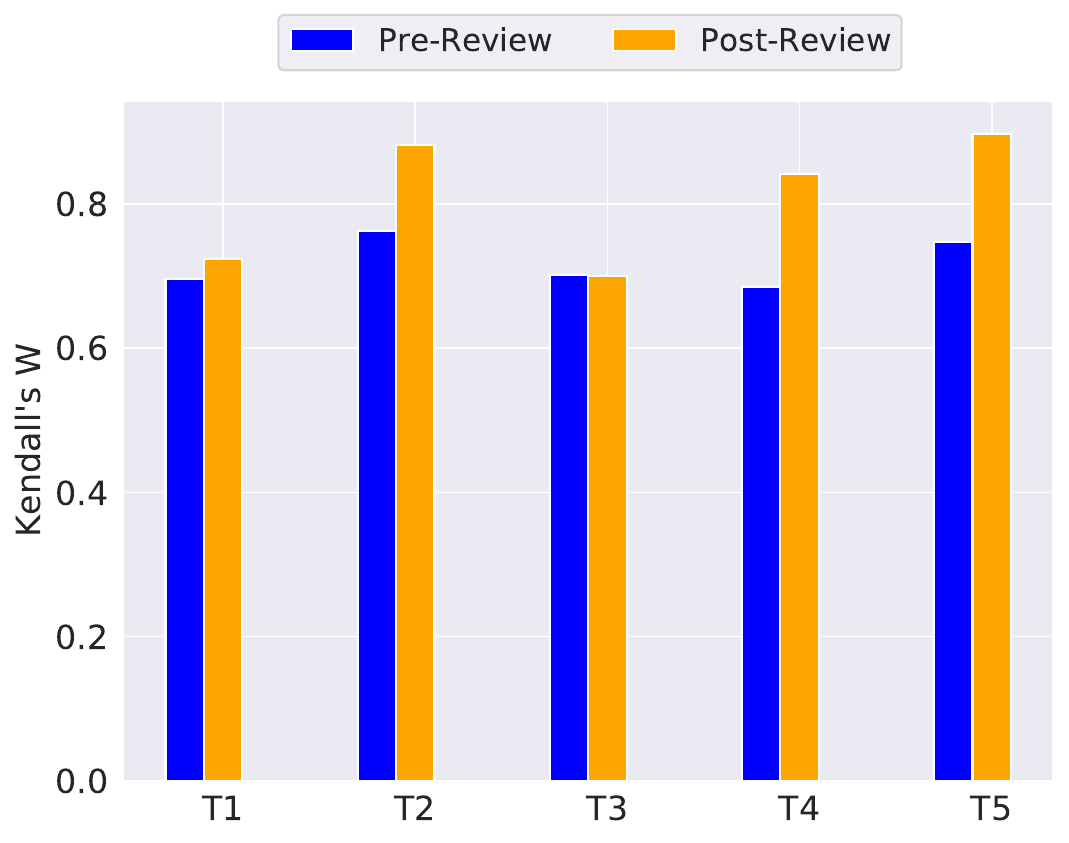}
    \caption{Kendall's W. The y-axis shows the values of Kendall's W statistic calculated on the respondents' implicit rankings. A higher value indicates a higher degree of consistency.}
    \label{fig:kendallsw}
    \end{subfigure}
    \hfill
    \caption{Effect of the interventions on the consistency of people's implicit relative judgments. The experimental conditions T1--T5 are shown on the x-axis.}
    \label{fig:conistency_metrics_pairwise}
\end{figure}

\subsubsection{Consistency of People's Implicit Relative Judgments}\hfill\\
\xhdr{Pairwise Consistency} 
We consider a measure of consistency analogous to the measure of accuracy described in Appendix \ref{subsec:appendix_accuracy}. We again derive people's implicit relative judgments from their absolute estimates. For each respondent and for each pair of apartments (A,B), we check if Apartment A was estimated to be more (>), less (<) or equally as expensive (=) as Apartment B. However, instead of comparing a respondent's implicit relative judgment with the ground truth ordering, we compare it to the other respondents' orderings---namely, to the majority vote of others' implicit relative judgments. 

In Figure~\ref{fig:majority_disagreement}, we show the average degree of disagreement between individual respondents' relative judgments and the majority vote. We observe that people's pre-review estimates are similarly consistent across all experimental conditions. In treatments T1 and T3, people's post-review estimates exhibit a similar degree of pairwise consistency as their pre-review estimates. However, in treatments T2, T4 and T5 we find that the average disagreement with the majority vote is decreased by 7.5, 6.9 and 7.1 percentage points respectively. It is important to note that the pre-review inconsistency was already quite low, leaving little room for improvement. The observed decrease in inconsistency in T2, T4 and T5 respectively correspond to 38\%, 32\% and 35\% of the total possible decrease. 

While this metric focused on pairwise consistency, below we consider a metric that quantifies the consistency in the overall ordering of the apartments: Kendall's W.

\xhdr{Kendall's W}
In order to assess the consistency of respondents' overall ordering of apartments, we treat the provided price estimates as implicitly ranking all of the apartments from the least expensive to the most expensive, allowing for ties. We then quantify the consistency between the respondents' implicit rankings utilizing Kendall's W, a non-parametric statistic for rank correlation.

Kendall's W is commonly employed to evaluate agreement amongst respondents in ranking tasks. At a high level, Kedall's W corresponds to the normalized sum of squared deviations from the mean in the rankings. We used the \texttt{Pingouin} library \cite{Vallat2018} to compute Kendall's W with a correction for ties. A value of one would indicate perfect agreement amongst respondents, while a value of zero would indicate no agreement.

In Figure~\ref{fig:kendallsw} we show the values of Kendall's W statistic calculated on the respondents' implicit pre-review and post-review rankings. Descriptively, we observe that in treatments T1 and T3, the respondents' pre-review and post-review rankings are consistent to a similar degree. However, treatments T2, T4 and T5 show an increase in Kendall's W of 0.12, 0.16 and 0.15 in the post-review rankings compared to the pre-review rankings.

\begin{figure}[t]
    \centering
    \includegraphics[width=0.48\textwidth]{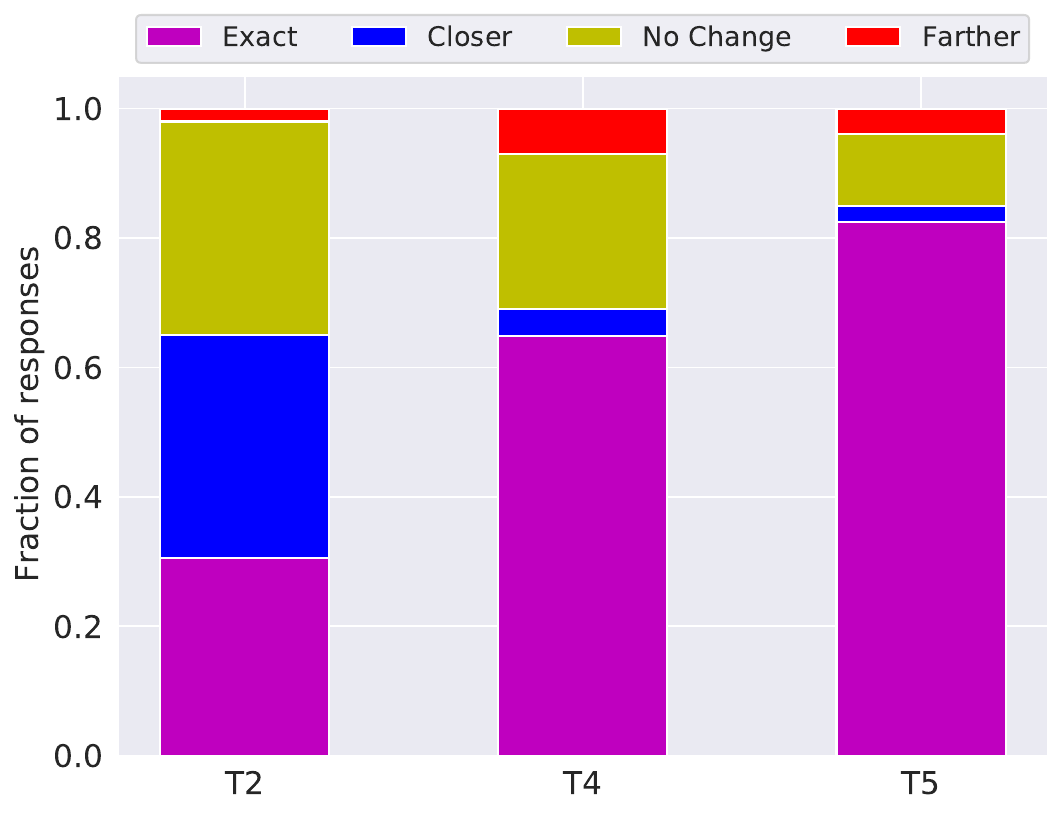}
    \caption{Change in agreement with machine predictions. The y-axis shows the fraction of pre-review estimates that were revised to match the decision aid's prediction exactly (Exact), that were revised in the direction of the prediction, but did not match the prediction exactly (Closer), that were not revised (No Change), and that were revised in the direction opposite of the prediction (Farther), for the experimental conditions T2, T4 and T5, shown on the x-axis.}
    \label{fig:dist_agreement_w_advice}
\end{figure}

\subsection{Agreement with Machine Advice} \label{subsec:appendix_advice}
In this section, we investigate the degree of agreement between the respondents’ estimates and machine advice. That is, we explore how observing machine advice impacted people's estimates. Did respondents adjust their initial estimates closer to machine predictions, or perhaps in the opposite direction? When respondents followed machine advice, did they copy it exactly or just move closer to it? We address these questions in Figure \ref{fig:dist_agreement_w_advice}.

In T2, respondents observed the decision aid's estimates of apartment prices. The majority of respondents' estimates were updated in the direction of machine advice (65\%). That is, the absolute difference between people's estimate and the decision aid's estimate became smaller for 65\% of responses. However, many estimates were not updated (33\%), and a few estimates were even updated in the direction opposite of machine advice (2\%). Amongst the 65\% of price estimates that were updated in the direction of the observed advice, 47\% were revised to exactly match the price predicted by the decision aid, while the remaining 53\% of estimates were just moved closer to the decision aid's predictions.

Unlike T2, where participants observed the decision aid's estimates of apartment prices, in T5 respondents observed only the decision aid's comparative valuation of pairs of apartments. Therefore, we cannot directly compare people's estimates and the decision aid's predictions. However, we can compare the decision aid's relative ordering of pairs of apartments and the respondents' implicit relative ordering based on their price estimates. As a running example, consider two apartments A and B, a decision aid that estimated A to be more expensive than B, and respondents who estimated A to be less or equally expensive as B. The price estimates of 85\% of pairs of apartments shown in the review phase were revised in the direction of machine advice (i.e., the respondents from the running example either increased their estimate for A, or decreased their estimate for B, or both). The implicit relative ordering of 11\% of pairs of apartments was not revised, and the estimates of 4\% of outlier pairs were revised in the direction opposite of machine advice (i.e., the respondents from the running example either decreased their estimate for A, or increased their estimate for B, or both). Among the 85\% of pairs that were revised in the direction of the decision aid's advice, 97\% of the price estimates were updated so that the apartments' implicit relative ordering matched the decision aid's relative ordering (i.e., the respondents from the running example revised their responses such that A is estimated to be more expensive than B), while for the remaining 3\% of pairs the price estimates were only moved closer to the decision aid's prediction (i.e., the respondents from the running example either increased their estimate for A, or decreased their estimate for B, or both, but A remained less or equally as expensive as B.)

In T4, participants did not observe any explicit machine advice. However, the pairs of apartments respondents were asked to review were selected by the decision aid that was also used in T5, thereby implicitly providing guidance. Hence, we again compare this decision aid's relative ordering with people's implicit relative ordering, for the pairs of apartments that were shown in the review phase. Despite not having a chance to observe the decision aid's relative ordering of apartments, many respondents updated their estimates in the direction of the decision aid's predictions. For 69\% of pairs estimates were revised in the direction of machine predictions, 24\% of implicit relative orderings were not revised, and only 7\% of estimates were revised in the direction opposite of machine predictions. Amongst the 69\% of pairs revised in the direction of the machine predictions, 94\% of the updates resulted in an implicit relative ordering that matched the decision aid's relative ordering (even though the respondents did not observe the decision aid's relative ordering), while for the remaining 6\% of pairs the price estimates were only moved closer to the decision aid's relative ordering.

\end{document}